\documentclass[12pt]{article}
\usepackage{tikz}
\usepackage{pgf}
\usepackage[colorlinks=true,linkcolor=black, citecolor=black,urlcolor=black]{hyperref}

\usepackage{multirow,graphics}
\usepackage{cite}
\usepackage{enumerate}
\usepackage{amstext}
\usepackage{amssymb}
\usepackage{amsmath}
\usepackage{booktabs}
\usepackage{graphicx}
\usepackage{subcaption}
\usepackage{color}
\usepackage{tikz}
\usepackage[nodisplayskipstretch]{setspace}
\usepackage{tcolorbox}
\usepackage{float}
\usepackage{wasysym}
\usepackage{ytableau}
\usepackage{textcomp}
\usepackage{blkarray}
\usepackage{color}
\usepackage{longtable}

\usetikzlibrary{decorations.pathmorphing}
\usetikzlibrary{decorations.markings}
\usetikzlibrary{intersections}
\usetikzlibrary{calc}
  \usetikzlibrary{positioning}
\usetikzlibrary{shapes}
\usetikzlibrary{fit}
\usetikzlibrary{backgrounds}

\tikzset{
photon/.style={decorate, decoration={snake}},
particle/.style={postaction={decorate},
    decoration={markings,mark=at position .5 with {\arrow{>}}}},
antiparticle/.style={postaction={decorate},
    decoration={markings,mark=at position .5 with {\arrow{<}}}},
gluon/.style={decorate, decoration={coil,amplitude=2pt, segment length=4pt},color=purple},
wilson/.style={color=blue, thick},
scalarZ/.style={postaction={decorate},decoration={markings, mark=at position .5 with{\arrow[scale=1]{stealth}}}},
scalarX/.style={postaction={decorate}, dashed, dash pattern = on 4pt off 2pt, dash phase = 2pt, decoration={markings, mark=at position .53 with{\arrow[scale=1]{stealth}}}},
scalarZw/.style={postaction={decorate},decoration={markings, mark=at position .75 with{\arrow[scale=1]{stealth}}}},
scalarXw/.style={postaction={decorate}, dashed, dash pattern = on 4pt off 2pt, dash phase = 2pt, decoration={markings, mark=at position .60 with{\arrow[scale=1]{stealth}}}},
frozen/.style={inner sep=0.7mm, rectangle,draw},
frozenblue/.style={rectangle, draw, fill=blue!20, inner sep=0.7mm},
blue/.style={rectangle, rounded corners, fill=blue!20, inner sep=0.7mm},
red/.style={rectangle, rounded corners, fill=red!20, inner sep=0.7mm},
>=stealth,
norm/.style={->, draw, shorten <=2pt, shorten >=2pt},
diag/.style={->, draw, shorten <=5pt, shorten >=3pt},
every node/.style={inner sep=0.5mm}
}

\newcommand{\ang}[1]
{
  \langle #1 \rangle
}

\newcommand{\cpthree}{$\mathbb{CP}^3$}
\newcommand{\Gr}[1]{$\operatorname{Gr}(#1)$}

\newcommand\be{\begin{equation}}
\newcommand\ee{\end{equation}}

\normalsize
\makeatletter
\divide\@tempdima\baselineskip
\@tempcnta=\@tempdima
\skip\@mpfootins = \skip\footins
\let\svthefootnote\thefootnote
\newcommand\blankfootnote[1]{%
  \let\thefootnote\relax\footnotetext{#1}%
  \let\thefootnote\svthefootnote%
}
\let\svfootnote\footnote
\renewcommand\footnote[2][?]{%
  \if\relax#1\relax%
    \blankfootnote{#2}%
  \else%
    \if?#1\svfootnote{#2}\else\svfootnote[#1]{#2}\fi%
  \fi
}
\newtheorem{conj}[equation]{Conjecture}

\definecolor{caribbeangreen}{rgb}{0.0, 0.8, 0.6}

\renewcommand{\@dotsep}{10000}
\makeatother

\begin{document}
\null\vskip-43pt \hfill
% \begin{minipage}[t]{30mm}
% \end{minipage}
\numberwithin{equation}{section}
\begin{center}
\phantom{vv}
\thispagestyle{empty}
\vspace{3cm}
\bigskip

{\Large \bf Cluster patterns in Landau and Leading Singularities via the Amplituhedron}

\bigskip
\mbox{
  \bf{\"Omer G\"urdo\u gan}$^1$, {\bf Matteo Parisi}$^1$
}
\footnote[]{ 
 $^1${\sffamily 
 \{\tt Omer.Gurdogan, Matteo.Parisi\}@maths.ox.ac.uk }
}
\bigskip

{\em $^1$\em Mathematical Institute, University of Oxford,\\
Andrew Wiles Building,
Woodstock Road,
Oxford,
OX2 6GG, United Kingdom.}

\vspace{3cm}
\bigskip
\vspace{30pt} {\bf Abstract}
\end{center} 

\noindent We advance the exploration of cluster-algebraic patterns in the
building blocks of scattering amplitudes in $\mathcal{N}=4$ super
Yang-Mills theory. In particular we conjecture that, given a maximal
cut of a loop amplitude, Landau singularities and poles of each
Yangian invariant appearing in any representation of the corresponding
Leading Singularities can be found together in a cluster. We check
these adjacencies for all one-loop amplitudes up to 9 points.  Along
the way, we also prove that all (rational) N$^2$MHV Yangian invariants
are cluster adjacent, confirming original conjectures.

\noindent

\newpage
\tableofcontents

\section{Introduction}
\label{sec:introduction}
Constructing scattering amplitudes from the knowledge of their their
singularities, i.e. their poles and branch-cut structure, is an approach
with a long history \cite{Eden:1966dnq}, which has proven to be
particularly effective for scattering amplitudes in $\mathcal{N}=4$ super Yang-Mills (SYM) theory.

Singularities of scattering amplitudes at tree-level are given by
multiparticle factorisation channels, which correspond to Mandelstam
invariants, and are constructed from subsets of the momenta of the
particles in the scattering process. Whereas, loop amplitudes exhibit
more complicated singularities, leading to logarithmic divergences. In
cases where loop amplitudes are expressed as (multiple)
polylogarithms, the collection of these logarithmic singularities is
called the symbol alphabet. When expressed in terms of momentum
twistors, many (all for $n\leq 7$, where $n$ is the number of
particles) of these are simply polynomials in the Pl\"ucker
coordinates in \Gr{4,n}. Moreover, their vanishing loci correspond to
special configurations of momentum twistors in \cpthree.

On one side, we have seen the emergence of positive geometries
\cite{Arkani-Hamed:2017tmz} as an overarching framework to geometrise
scattering ampltiudes and their analytic structure, at tree-level and
for loop integrands in several theories, among which $\mathcal{N}=4$
SYM.  In 2013 a full geometric description for tree-level and
integrands of loop-level scattering amplitudes in planar
$\mathcal{N}=4$ SYM has been proposed in \cite{Arkani-Hamed:2013jha},
under the name of \emph{amplituhedron}. Others constructions followed
few years later \cite{He:2018okq,Damgaard:2019ztj}.

On the other, we have witnessed an increasing appearance of cluster
algebra structures in scattering amplitudes, especially in capturing
singularities of (integrated) loop amplitudes in $\mathcal{N}=4$ SYM.
This started in 2013 with the conjecture made by Golden et al in
\cite{Golden:2013xva} that the symbol letters of six and-
seven-particle loop amplitudes are $\mathcal{A}$-coordinates the
\Gr{4,n} cluster algebra.  Few years later in \cite{Drummond:2017ssj}
it was conjectured that these letters satisfy remarkable cluster
properties, called \emph{cluster adjacency}. In terms of the symbol,
they dictate which letters can appear consecutively.  Moreover,
shortly after these adjacencies were observed at tree-level as well by
themselves, and in connection with symbol entries
\cite{Drummond:2018dfd}, (see also \cite{MSSV2020} for a recent work
on the cluster-adjacency of one-loop amplitudes). The guidance of
cluster algebras has unlocked the possibility of developing a powerful
bootstrap programme which allowed to perform computations that
otherwise would have been beyond reach \cite{Dixon:2011pw,
  Dixon:2011nj, Dixon:2013eka, Dixon:2014voa,
  Dixon:2014iba,Drummond:2014ffa,Dixon:2015iva,Caron-Huot:2016owq,Caron-Huot:2019vjl,Drummond:2018caf}. At
the same time, they shed more light on the mathematical structures
describing singularities of scattering amplitudes and motivate the
existence of a possible geometric origin.

One manifestattion of the cluster-algebraic phenomena is an
observation that building blocks of a BCFW representation of the
tree-level amplitude, which are \emph{Yangian invariants}, are
\emph{cluster adjacent} \cite{Drummond:2018dfd}. In other words, all
poles of each of them are expressed by a collection of
$\mathcal{A}$-coordinates of the $\operatorname{Gr}(4,n)$ cluster
algebra that can be found together in common cluster. Moreover, this
conjecture was generalised in \cite{Mago:2019waa}, for all (rational)
Yangian invariants of $\mathcal{N}=4$ SYM.  In geometric
terms, poles of (rational) Yangian invariants are codimension-one
boundaries of the so-called \emph{generalised triangles} of the
amplituhedron \cite{Lukowski:2020dpn,Lukowski:2019sxw}. Furthermore,
different representations of scattering amplitudes, obtained from
identities among Yangian invariants, correspond to different
\emph{triangulations} of the same geometric space, i.e. the
amplituhedron.

One of the first steps towards an amplituhedronic understanding of
cluster phenomena was taken in \cite{Lukowski:2019sxw}, where a toy
model for tree-level cluster adjacency of $\mathcal{N}=4$ SYM was
considered. It was proved that Yangian invariants of the $m=2$
amplituhedron are cluster adjacent with respect to the well known
$\operatorname{Gr}(2,n) \simeq A_{n-3}$ cluster algebra.  The $m=2$
amplituhedron is often considered as a toy-model for the physical
$m = 4$ case, moreover it also governs the geometry of one-loop MHV
integrands \cite{Arkani-Hamed:2014dca} and it has some relevance for
the NMHV ones as well \cite{Kojima:2020tjf}.  By exploiting the geometry
of the $m=2$ amplituhedron, an explicit
expression of all Yangian invariants was provided in \cite{Lukowski:2019sxw}, where cluster
adjacency of their poles is manifest.

The interest in understanding how cluster algebras encode the analytic
properties of scattering amplitudes led physicists to explore the
connection between cluster algebras and the \emph{positive tropical
  Grassmannian}, originally introduced in \cite{Speyer2003TheTT}.  See
for examples
\cite{Drummond:2019cxm,Henke:2019hve,Arkani-Hamed:2019rds}, for
applications in $\mathcal{N}=4$ SYM.  Remarkably, the very same
positive tropical Grassmannian has been found to regulate the
combinatorics of \emph{triangulations} (and, more generally,
subdivisions) of the $m=2$ amplituhedron \cite{Lukowski:2020dpn}. This
raises the question on whether there is a deeper connection between
the latter object and cluster algebras themselves.

A remarkable instance of how geometry encodes singularities of
scattering amplitudes in $\mathcal{N}=4$ SYM is the fact that all
\emph{Leading Singularities} of the theory, at any loop order, can be
computed by a contour integral over the space of $k$-planes in $n$
dimensions, called \emph{Grassmannian}
\cite{ArkaniHamed:2009dn,Mason:2009qx}.  Leading Singularities are the
singularities of the integrand of a loop amplitude with maximal
codimension in loop momenta.  The geometrisation has been pushed even
further via \cite{ArkaniHamed:2012nw} and, a year after, the authors
of \cite{Arkani-Hamed:2013jha} defined the \emph{loop amplituhedron},
whose boundaries encode singularities of the integrand, among which
are the Leading Singularities corresponding to maximal cuts.

The application of this geometric approach to \emph{Landau
  Singularities} \cite{Dennen:2016mdk, Prlina:2017tvx, Prlina:2017azl}
is another example of its utility to obtain a better understanding of
the structure of singularities of scattering amplitudes. The Landau
analysis allows to connect singularities of the \emph{integrand},
described geometrically from boundaries of loop ampliuthedra, to the
ones of the \emph{integrated} amplitudes.  Among all Landau
singularities, there are in general many spurious ones coming from
summing over Feynmann diagrams.  On the other hand, the amplituhedron
can tell which are the true singularities of the integrand, and
therefore select the true Landau singularities of the loop amplitude.

In this work, using an amplituhedron-based approach to encode building blocks of scattering amplitudes in $\mathcal{N}=4$ SYM, i.e. Yangian invariants and Leading Singularities, we explore cluster patterns between the latter and Landau Singularities.

We first review the preliminary concepts appearing in our work in
Section \ref{sec:review}. In particular, we review the notion of
\emph{cluster adjacency} in \ref{sec:cluster-adjacency} and state its
known various incarnations; then in Section \ref{sec:LeSAH} we
introduce the concepts of \emph{Leading Singularities} and the
\emph{(loop) amplituhedron}, and how one can obtain the former from
special boundaries of the latter; in Section \ref{sec:LaS} we present
the definition of \emph{Landau singularities} and how the loop
amplituhedron can select the non spurious ones; for both Leading and
Landau singularities we present in the respective sections examples at
one-loop which will be relevant for our work.

In Section \ref{sec:clust-adjac-yang} we will prove \emph{cluster adjacency for all (rational) N$^2$MHV Yangian invariants}. In particular, we introduce the geometric method used to determine the actual poles of Yangian invariants in terms of cluster variables, and we present the results in Section \ref{sec:adjac-all-rati}; finally, in Section \ref{sec:four-mass-box} we prove that Yangian invariants of the four-mass box type violate cluster adjacency.

In Section \ref{sec:cluster-adjacency-at-one-loop} we present the main
conjecture of our paper: \emph{cluster adjacency between Leading and
  Landau singularities}, which we abbrehivate as ``LL-cluster
adjacency''. We first introduce how to find all Yangian invariants
which can be used to represent a given Leading Singularity from the
geometry of the loop amplituhedron. We then present our checks and
proofs about these adjacencies in Sections
\ref{sec:LLNMHV},\ref{sec:LLN2MHV} for all one-loop amplitudes up to 9
points for the NMHV and N$^2$MHV cases, respectively. In Section
\ref{sec:NMHV7explicit} we show the one-loop NMHV 7 points amplitude
in a representation which is uniquely fixed by LL cluster adjacency.
Finally, in Section \ref{sec:conl} we end with conclusions and
directions for future works.

\section{Cluster algebras, Singularities and Geometry}\label{sec:review}
\subsection{Cluster adjacency}
\label{sec:cluster-adjacency}

We begin by reviewing  the notion of cluster adjacency for singularities of
scattering amplitudes in planar $\mathcal{N}=4$ SYM. These are observations about the
appearance of singularities of the amplitudes in relation to how they are
encoded in a corresponding \Gr{4,n} cluster algebra.

The mathematics literature on cluster algebras,
e.g. \cite{1021.16017,1054.17024,SCOTT_2006}, provides an excellent
introduction to the concept.  Aspects of cluster algebras of
Grassmannian-type in the context of scattering amplitudes have been
explained in detail in \cite{Golden:2013xva,
  Drummond:2017ssj,Drummond:2018dfd}. We will therefore introduce only
the cluster-terminology which will be employed in stating our results.

One way of representing clusters of \Gr{4,n} cluster algebras are
quiver diagrams. These have $3(n-5)$ distinct nodes, called
\emph{$\mathcal{A}$-coordinates}, that are in general polynomials in the
Pl\"ucker coordinates of \Gr{4,n}. When $n$ is greater than $8$, there are
infinitely-many clusters and therefore infinitely-many $\mathcal{A}$-coordinates. Remarkably, all known \emph{rational} singularities of BDS-like
normalised amplitudes are $\mathcal{A}$-coordinates of
\Gr{4,n} cluster algebras \cite{Golden:2013xva}.

Each cluster in a given cluster algebra can be obtained from any other cluster by (sequences of)
\emph{mutations}. A mutation, expressed in terms of $\mathcal{A}$-coordinates, is an operation which
replaces a chosen node of the quiver with a new value, as well as
locally changing the connectivity of the quiver diagram. Two
$\mathcal{A}$-coordinates are said to be \emph{cluster adjacent} if there exists a cluster in which they appear together.

As a toy model, one can consider \Gr{2,n} cluster algebra where
clusters correspond to triangulations of an $n$-gon and the
$\mathcal{A}$-coordinates correspond to the chords of this
triangulation. Mutations act as flipping the chord inside the
quadrilateral that they are the diagonal of. In this case,
cluster-adjacent coordinates correspond to non-crossing chords. Two
coordinates are not cluster-adjacent if and only if the
corresponding chords cross, i.e. they mutate to each other.

\Gr{4,n} cluster algebras are more complicated and allows for other
adjacency situations. In particular, pairs of \Gr{4,n}
$\mathcal{A}$-coordinates can never appear in a cluster together even
though there is no mutation that relates to them. Therefore there is
no known simple geometric picture from which one can infer
(collective) adjacencies of sets of these variables.\\

\noindent There are various different but related cluster-adjacency
statements for scattering amplitudes in $\mathcal{N}=4$ super
Yang-Mills:
\paragraph{Adjacency of symbol letters:} Two $\mathcal{A}$-coordinates
appear next to each other in the symbol of a BDS-like normalised
amplitude only if there is a cluster that contains both of them
\cite{Drummond:2017ssj}. For all known integrable words with physical
initial entries, this requirement appears to be equivalent to extended
Steinmann conditions of \cite{PapathanasiouAmps17,Caron-Huot:2019bsq}.

\paragraph{Poles of tree BCFW amplitudes} In
\cite{Drummond:2018dfd} it is conjectured that BCFW representations of
tree amplitudes in $\mathcal{N}=4$ SYM are linear combinations of terms whose poles
are mutually cluster adjacent in a strict sense. Moreover it was
observed in several examples that it is possible find a cluster in the
relevant cluster algebra which contains all poles of each BCFW term.

The simplest case of this statement is for NMHV tree amplitudes\footnote{We will denote tree-level N$^2$MHV $n$-points amplitudes as $\mathcal{A}_{n,k}$.}, that
are sums of $R$-invariants:
\begin{equation}
  \mathcal{A}_{n,1}
  =
  \sum_{1<i<j<n} R_{1\, i\, i+1\,j+1}\, ,
\end{equation}
and the adjacency for the latter has been proven in
\cite{Drummond:2018dfd} through a procedure in which one starts from
the initial cluster of \Gr{4,6} and arrives at a cluster containing
the poles of $R_{1\, i\, i+1\,j+1}$ through a sequence of
(partial) cyclic rotations.

This observation, in particular the simple proof of the
cluster-adjacency of $R$-invariants, motivates the question of how far
this property extends. 
In \cite{Mago:2019waa} it was conjectured that all (rational) Yangian invariants satisfy such cluster adjacent properties.
It is also natural to ask whether this is a
mathematical property of Yangian invariants or whether it is an extra
physical constraint that BCFW terms are expected to satisfy.

\paragraph{Rational Yangian invariants}
The natural question of whether the manifestation of cluster adjacency
in BCFW terms extends to more general Yangian Invariants was asked in
\cite{Drummond:2018dfd} and affirmative evindence was given for
rational Yangian invariants in \cite{Mago:2019waa} through an argument
via the Sklyanin bracket, along with a conjecture that this should
hold for all such Yangian invariants.

\paragraph{R-invariants and NMHV final entries}
Finally, the fourth statement of cluster adjacency concerns NMHV loop
amplitudes, which are sums of iterated integrals whose coefficients
are R-invariants. Schematically they have the form
\begin{equation}
  \mathcal{A}^{(L)}_{n,1}
  =
  \sum_{\alpha,i_1,\dotsc,i_{2L}}\,
  R_{\alpha}\,c_{1,\dotsc,2L}\,\, \phi_{i_1} \otimes \dotsm \otimes \phi_{i_{2L}},
\end{equation}
where the index $\alpha$ enumerates all relevant $R$-invariants, $L$ is the loop order, and
the indices $i_k$ enumerate letters, the rational ones of which are
$\mathcal{A}$-coordinates of the \Gr{4,n} cluster algebra.

The observation which holds for all known such amplitudes is that the
final entries of the symbols of the polylogarithms multiplying the
$R$-invariants are cluster-adjacent to all of the poles of the
R-invariant that multiplies them.

It is also worth remembering that in general one needs to write out
the amplitude with a redundant set of $R$-invariants that satisfy
linear 6-term identities in order to make manifest this
cluster-adjacency property.

\subsection{Leading Singularities from the amplituhedron}\label{sec:LeSAH}
We review here the concept of Leading Singularities. In particular, we show how Leading Singularities for $\mathcal{N}=4$ SYM can be computed more geometrically via a Grassmannian approach, and via the loop amplituhedron.

\paragraph{Leading Singularities.}
The concept of leading singularities was originally introduced within the \emph{Analytic Bootstrap Programme} in the 1960’s \cite{Eden:1966dnq}. At the beginning of this century, with the advent of novel on-shell techniques such as \emph{generalised unitarity}, the concept of Leading Singularities has been broadly employed and exploited in computation of scattering amplitudes, in particular in Yang-Mills \cite{Britto:2004nc}.  

Loop amplitudes in planar $\mathcal{N}=4$ SYM are computed from \emph{integrands}, which are rational functions of external kinematics and loop momenta, by integration over particular real-contours in the $4L$-dimensional loop momentum space. However, in general this contour is known not to preserve the symmetries of the theory, and leads for example to IR-divergences.
In this regards, it might seems natural to choose complex contours corresponding to computing residues of the integrand.
Leading singularities are then the residues of the integrand computed around tori encircling the loci where a maximal set of internal propagators (e.g. four for one-loop) go on-shell.

Given a 1-loop $n$-points scattering amplitude $\mathcal{A}(1,\ldots,n)$ and a partition $\mathcal{C}$ of $\lbrace 1,\ldots,n \rbrace$ into 4 disjoint subsets $I_1,\ldots,I_4$, then the \emph{Leading Singularity} of the amplitude is defined as:
\begin{equation}
\int \prod_{a=1}^4 d^4 \eta_a \, d^4 \ell_a \, \delta(\ell^2_a) \, \prod_{a=1}^4 \, \mathcal{A}_a \left( \lbrace \ell_a,\eta_a \rbrace, I_a, \lbrace -\ell_{a+1},\eta_{a+1} \rbrace \right) 
\end{equation}
where the index $a$ is mod 4, the integral over $\ell$ is localised over the solutions of the delta function and the integral over the Grassmann coordinates\footnote{See \cite{ArkaniHamed:2008gz} for a good review on $\mathcal{N}=4$ SYM and its conventions.} $\eta_a$ amounts to sum over all possible internal states flowing between the different sub-amplitudes $\lbrace \mathcal{A}_1, \ldots \mathcal{A}_4\rbrace$. 
Since the four internal propagators are forced to vanish by the delta function, the internal particles can be taken on-shell. Therefore leading singularities are in general
simply the products of tree-amplitudes, summed over all the internal particles which can be exchanged, and integrated over the on-shell phase space of each.

\paragraph{Leading Singularities from the Grassmannian.}
In \cite{ArkaniHamed:2009dn}, leading singularities were proposed as the complete set of IR-finite quantities that contains all the information needed to compute the S-Matrix of $\mathcal{N}=4$ SYM. 
Beautifully, both in momentum space and in momentum twistor space \cite{Mason:2009qx}, all leading singularities of the theory, at any loop order, can be computed by a contour integral over the space of $k$-planes in $n$ dimensions, called \emph{Grassmannian} and denoted as $\operatorname{Gr}(k,n)$. Here $k$ is the helicity sector of the amplitude.
 Remarkably, in \cite{ArkaniHamed:2012nw}, it was shown that only the `positive' part $\operatorname{Gr}_+(k,n)$ of this space, called the \emph{positive Grassmannian} \cite{lusztig,postnikov}, is relevant for scattering amplitudes. Moreover, the integration contour providing Leading Singularities is performed on some of its \emph{positroid cells}, in terms of which the positive Grassmannian has a beautiful stratification.
 
 All positroid cells are in bijection with various nice combinatorial objects, including equivalence classes of reduced plabic graphs, also known as \emph{on-shell diagrams} in the context of scattering amplitudes.
A comprehensive summary about on-shell diagrams, their classification, evaluation, and relations, has been described  in \cite{ArkaniHamed:2012nw}. 
Formulae for 1-loop Leading Singularities for $\mathcal{N}=4$ SYM in momentum twistor variables are reported in \cite{Bourjaily:2013mma} using on-shell diagrams, and will be used in an example in Section \ref{sec:LeS1Loop}.

\paragraph{The loop amplituhedron.}
In 2013, the emergence of scattering amplitudes from polytopes \`a la
Hodge \cite{Hodges:2009hk} and from the positive Grassmannians, came
together in Arkani-Hamed and Trnka's work \cite{Arkani-Hamed:2013jha}.
They introduced a novel mathematical object called the
amplituhedron. Its canonical form gives all tree-level and the
integrand of loop-level scattering amplitudes in planar
$\mathcal{N}=4$ Yang-Mils. Its boundaries geometrically encode all the
singularities of the latter. 

Let us fix a totally positive matrix
$Z \in \mbox{M}_+(n,k+4)$, i.e. all its maximal minors are positve.
 Let us consider $Y \in Gr(k,k+4)$ and $L$
lines $\mathcal{L}^{(l)} \in Gr(2,4+k), l=1,\ldots,L$, called
\emph{loop momenta}, in the four-dimensional complement of $Y$.  Then
the \emph{loop amplituhedron} $\mathcal{A}^{(L)}_{n,k}$ is the set\footnote{With abuse of notation, we will denote $\mathcal{A}^{(L)}_{n,k}$ as both the amplituhedron and the corresponding amplitude. It will be clear from the context which one we will be referring to.} of
$(Y,\mathcal{L}^{(1)},\ldots,\mathcal{L}^{(L)})$ such that:
\begin{equation}\label{def:loopAH}
Y=C \cdot Z, \quad \mathcal{L}^{(l)} =D^{(l)}\cdot Z, \quad l=1,\ldots,L
\end{equation}
where $C \in Gr(k,n)$, $D^{(l)} \in Gr(2,n)$ is in the four-dimensional complement of $C$ and such that all $(k+2s)\times n$ matrices of the form
\begin{equation}\label{eq:poscond}
\begin{pmatrix}
C \\
D^{(i_1)} \\
\vdots \\
D^{(i_s)}
\end{pmatrix}
\end{equation}
are \emph{totally positive}, i.e. they have all their maximal minors positive, with $0\leq s \leq L$.

An alternative definition of the loop amplituhedron based on \emph{sign flips} and inequalities was introduced in \cite{Arkani-Hamed:2017vfh}, and conjectured to be equivalent to the definition given above. 
The canonical form of the loop amplituhedron $\mathcal{A}^{(L)}_{n,k}$ encodes the integrand of the $L$-loop N$^k$MHV $n$-points amplitude.

Let now introduce some notation which will be useful in the following. Let us denote determinants of the $(4+k) \times (4+k)$ matrices obtained by stacking together $Y$ and rows of $Z$ specified by the indices $i,j,l,s$ as:
\begin{equation}\label{eq:brackets}
\langle Y i \, j \, l \, s \rangle :=\epsilon_{A_1 \ldots A_{k}BCDE} Y^{A_1}_1 \ldots Y^{A_k}_k Z_i^{B}Z_j^{C}Z_l^{D}Z_s^E,
\end{equation}
and analogously for brackets of the type $\langle Y \mathcal{L}^{(l)} i \, j \rangle$. 
As explained in \cite{Arkani-Hamed:2017vfh}, one can go from the space of bosonised momentum twistors where $\mathcal{L}^{(l)}$ and $Z_i$ live to the space of physical momentum twistors\footnote{For conventions on momentum twistors, which we will denote as $z_i$, in a similar context see e.g. \cite{Mago:2019waa}.} in $\mathbb{P}^3$ by projecting them through $Y$. Therefore one can identify the following brackets:
\begin{equation}
\langle a \, b \, c \, d \rangle \equiv \langle Y a \, b \, c \, d \rangle
\end{equation}
where the left hand side are brackets in momentum twistors and the right hand side are brackets defined in \eqref{eq:brackets}. In the following, with abuse of notation, we will sometime denote both cases as $\langle a \, b \, c \, d \rangle$ and it will be clear from the context.

\paragraph{Yangian Invariants from the amplituhedron}
For $L=0$, i.e. tree-level, the definition of loop amplituhedron in \eqref{def:loopAH} reproduces the definition of the  \emph{tree amplituhedron} \cite{Arkani-Hamed:2013jha}. This is the set of $Y \in Gr(k,k+4)$ such that:
\begin{equation}\label{def:AH}
Y=C \cdot Z,
\end{equation}
where $C \in Gr_+(k,n)$ and $Z$ is the fixed totally positive $n \times (k+4)$ matrix defined above. 
The tree amplituhedron is therefore the image of the map $\tilde{Z}$ induced by the fixed matrix $Z$, from the positive Grassmannian $\operatorname{Gr}_+(k,n)$ to another Grassmannian $\operatorname{Gr}(k,k+4)$, i.e. $\tilde{Z}: C \mapsto C \cdot Z =Y$. This map is not injective, since the dimension of the amplituhedron is $4k$, whereas the dimension of $\operatorname{Gr}_+(k,n)$ is $k(n-k)$, which is in general higher. 

Let us consider  positroid cells in $\operatorname{Gr}_+(k,n)$ which have the same dimension as the amplituhedron, i.e. $4k$, and have a full dimensional image into the amplituhedron. This is equivalent to considering $4k$-dimensional cells which, within the physics community, are said to have `kinematic support', and were studied and classified in the context of the positive Grassmannians \cite{ArkaniHamed:2012nw}. These cells are the same appearing as integration contours of the Grassmannian integral formulae in momentum twistors \cite{Mason:2009qx}. 
Given a $4k$-dimensional cell $S$ in $\operatorname{Gr}_+(k,n)$ with kinematic support, if we perform such integral over $S$ we obtain a building block $\mathcal{Y}_S$ for Leading Singularities of $\mathcal{N}=4$ SYM (hence all tree-level amplitudes as well), which is referred to as\footnote{In \cite{Drummond:2010qh}, it was indeed shown that the integral enjoys an infinite dimensional symmetry, which is the Yangian of $psu(2,2|2)$, and is simply called \emph{the Yangian} in literature on scattering amplitudes. This symmetry is the hallmark of integrability of $\mathcal{N}=4$ SYM.} N$^k$MHV $n$-particle \emph{Yangian invariant}. 
In this work, with a slight abuse of terminology, we will also refer to the image of $S$ into the amplituhedron as \emph{Yangian invariant}, and denote it as $Y_S$. 
It will be clear from the context which one we will be referring to.

For example, for NMHV amplitudes Yangian invariants are called \emph{R-invariants}, which can be compactly expressed as
\begin{equation}
R_{12345}=\frac{\delta^{0|4}(\langle 1234 \rangle \chi_5 + \mbox{cyclic})}{\langle 1234 \rangle \langle 2345 \rangle \langle 1345 \rangle \langle 1245 \rangle \langle 1235 \rangle},
\end{equation}
where the brackets are in momentum twistors are simply $\ang{ijkl}=\epsilon_{ABCD}z^A_i z^B_j z^C_k z^D_l$, and $\chi_i$ are the Grassmann variables used to express amplitudes in $\mathcal{N}=4$ SYM in super-momentum twistors. Analogously, we will denote as $R_I$ the analogous R-invariant with momentum twistor listed in $I$.

Among $4k$-dimensional cells with kinematic support, corresponding to Yangian invariants, there are cells which are mapped injectively into the amplituhedron. 
The image of such cells into the amplituhedron are referred to as \emph{generalised triangles} in \cite{Lukowski:2020dpn,Lukowski:2019sxw}. Generalised triangles are elements of triangulations of the amplituhedron. 
Cells corresponding to generalised triangles have  \emph{intersection number} one, see \cite{ArkaniHamed:2012nw} for more details, and their corresponding Yangian invariants\footnote{Yangian invariants of this type were called \emph{rational} in \cite{Mago:2019waa}.} are the building blocks for tree-level scattering amplitudes.
Whereas, $4k$-dimensional cells with kinematic support, but with intersection number \emph{higher} than one, are not mapped injectively in the amplituhedron: points in the image have a finite number (bigger than one) of pre-images in the cell.
Yangian invariants associated with this type of cells do not enter representations of scattering amplitudes at tree-level, but are relevant for their Leading Singularities. These Yangian invariants can be written as a sum of terms which in general are algebraic (e.g. contain square-roots), but the sum is still rational. See Section \ref{sec:four-mass-box} for a relevant example.

\paragraph{Leading Singularities from the amplituhedron.}
Let us now consider the boundaries of the loop amplituhedron and understand how these are related to Leading Singularities. 
It is known that the boundaries of the tree amplituhedron are on the vanishing locus of:
\begin{equation} \label{eq:treesing}
\langle Y i_1 \, i_1+1 \, j_1 \, j_1+1 \rangle=0, \ldots, \langle Y i_d \, i_d+1 \, j_d \, j_d+1 \rangle=0
\end{equation}
for some $d>0$ and all indices (considered cyclically) in $\lbrace 1, \ldots, n \rbrace$. In order to make connection with Leading Singularities, we will not focus on this tree-level type of boundaries. Instead, we will consider boundaries where $\mathcal{L}^{(l)}$ satisfies any of the following conditions, called \emph{on-shell conditions}:
\begin{equation}\label{eq:onshellcond}
\langle Y \mathcal{L}^{(l_1)} i_1 j_1 \rangle=0, \ldots, \langle Y \mathcal{L}^{(l_d)} i_d j_d \rangle=0, \, \langle Y \mathcal{L}^{(s_1)} \mathcal{L}^{(s_2)} \rangle=0
\end{equation}
for some $l_a,s_1,s_2 \in \lbrace 1, \ldots, L \rbrace$ and $i_a \in \lbrace 1, \ldots, n \rbrace$, and $Y$ does not lie on any of the tree-level type boundaries in equation  \eqref{eq:treesing}. 
Each set $\mathcal{C}$ of on-shell conditions has a certain number of solutions $\lbrace \mathcal{L}^*_a \rbrace_{\mathcal{C}}$, where we denoted as $\mathcal{L}^*_a$ the corresponding collection of L lines $({\mathcal{L}^*}^{(1)},\ldots,{\mathcal{L}^*}^{(L)})$. Following the terminology of \cite{Prlina:2017azl}, boundaries of the type \eqref{eq:onshellcond} are called \emph{$\mathcal{L}$-boundaries} and the connected components associated to each different solution of the same on-shell condition are called \emph{branches}.
If it exists, we denote as $\mathcal{B}[\mathcal{C},\mathcal{L}^*]$ the boundaries of the loop amplituhedron, which are $\mathcal{L}$-boundaries determined by the set of on-shell conditions $\mathcal{C}$ and are in the  branch corresponding to the solution $\mathcal{L}^*$. 
In \cite{Prlina:2017azl}, it has been showed that, once we fix $\mathcal{C}$ and $\mathcal{L}^*$, there exists a minimum\footnote{Using \emph{parity}, which is a symmetry of scattering amplitudes and of the amplituhedron, one can also establish an upper bound as: $k \leq n- \bar{k}_{min}-4$, where $\bar{k}_{min}$ is the minimal value of $k$ for which the parity-conjugated branch appears.} $k_{min}$ such that the loop amplituhedron $\mathcal{A}^{(L)}_{n,k}$ has the boundaries $\mathcal{B}[\mathcal{C},\mathcal{L}^*]$ for all $k \geq k_{min}$.

Finally, we will focus on the $\mathcal{L}$-boundaries which are relevant for Leading Singularities, which corresponds to \emph{maximal cuts}. If $\mathcal{C}$ is a set of on-shell conditions, then $\mathcal{C}$ is a \emph{maximal-cut} if it is maximal by inclusion, i.e. we can not add more on-shell conditions to $\mathcal{C}$ with $Y$ not being on tree-level type boundaries of equation  \eqref{eq:treesing}. In particular, an $\mathcal{L}$-boundary associated to a maximal-cut has codimension $4L$ and the solutions in each branch have loop momenta localised in points $\lbrace \mathcal{L}^* \rbrace$.

For a maximal cut $\mathcal{C}$, boundaries $\mathcal{B}[\mathcal{C},\mathcal{L}^*]$ of the loop amplituhedron correspond to Leading Singularities of the amplitude $\mathcal{A}_{n,k}^{(L)}$.
In particular, as one can extract tree-level amplitudes $\mathcal{A}_{n,k}$ from the canonical form of the tree amplituhedron, one can extract the Leading Singularities $\mathrm{LeS}[\mathcal{C},\mathcal{L}^*]$ from the canonical form of the codimension-$4L$ boundaries $\mathcal{B}[\mathcal{C},\mathcal{L}^*]$ of the loop amplituhedron .

It is known that all Leading Singularities of an amplitude $\mathcal{A}_{n,k}^{(L)}$ can be expressed as a sum of $n$-particles $N^kMHV$ Yangian invariants, i.e. for a certain Leading Singularity $\mathrm{LeS}$ there is a collection of $4k$-dimensional cells $\lbrace S_a \rbrace$ in $\operatorname{Gr}_+(k,n)$ with kinematic support, such that:
\begin{equation}\label{eq:LeSSumYI}
\mathrm{LeS}=\sum_{a} \mathcal{Y}_{S_a}.
\end{equation}
This is just a rephrasing of the conjecture that the Grassmannian integral representation of scattering amplitudes provides Leading Singularities if integrated over proper contours, such as the one\footnote{With suited orientation of each cell.} provided by the above collection of cells $\lbrace S_a \rbrace$.
The sum in \eqref{eq:LeSSumYI} is the geometrical equivalent of `triangulating' the boundary of the loop amplituhedron, corresponding to the Leading Singularity, with the collection of Yangian invariants $\lbrace Y_{S_a}\rbrace$.
As different representations of a scattering amplitude $\mathcal{A}_{n,k}$ (tree-level or loop integrand) are just different ways to triangulate the amplituedron (tree or loop), different representations of a Leading Singularity $\mathrm{LeS}$ as sum of Yangian invariants correspond to different triangulations of corresponding boundary of the loop amplituhedron.

In Section \ref{sec:cluster-adjacency-at-one-loop}, we will exploit the geometric definition of the loop amplituhedron to compute \emph{all} Yangian invariants which can be part of a triangulation of a given boundary of the loop amplituhedron, i.e. all Yangian invariants which can be used to express a given Leading Singularity. Moreover, we will see how this connects to the Landau analysis in the next section. 

\subsubsection{Leading Singularities at 1 loop}\label{sec:LeS1Loop}
In this section, we will provide an illustrative example on how to compute Leading Singularities from the Grassmannian for one-loop NMHV, following \cite{Bourjaily:2013mma} (in particular, see Table 3). We will consider only some maximal-cuts which will be relevant for our analysis. We will briefly comment on the N$^2$MHV case, and we will employ a different strategy based on the amplituhedron described in Section \eqref{sec:clust-adjac-yang}.

Given a cut $\mathcal{C}=\lbrace I_1,\ldots,I_4 \rbrace$ for the loop amplitude $\mathcal{A}^{(1)}_{n,k}$, on-shell diagrams with tree sub-amplitudes $\mathcal{A}_{n_1,k_1}(I_1) \otimes \ldots \otimes \mathcal{A}_{n_4,k_4}(I_4)$, such that:
\begin{equation}\label{eq:relhelicity}
\sum_{a=1}^4 k_a=k-2, \quad \sum_{a=1}^4 n_a=n+8,
\end{equation}
correspond to Leading Singularities of $\mathcal{A}^{(L=1)}_{n,k}$.
Here we denoted sub-amplitudes as $\mathcal{A}_{n',k'}(I')$, where $k'$, with $0\leq k'\leq n-4$, is its N$^{k'}$MHV helicity sector\footnote{For $n=3$, we also admit $k'=-1$, which corresponds to $\overline{\mathrm{MHV}}$, i.e. a \emph{white} vertex. Moreover, note that $\mathcal{A}_{0}(\ldots)=1$, since we are in the momentum twistor space.}, $n'$ the number of legs, and $I'$ denotes the indices the external particles contained. 
\paragraph{Leading Singularities for NMHV 1 loop.}
Let us now list the types of Leading Singularities which can appear at NMHV at one-loop. By equation \eqref{eq:relhelicity} we must have:
\begin{equation}
k_1+k_2+k_3+k_4=1-2=-1 
\end{equation}
Since we can have $k_a=-1$ only when one of the sub-amplitude is a 3 point amplitude, otherwise $k_a$ are positive, then we must have at least a 3-point subamplitude to satisfy equation  \eqref{eq:relhelicity}. Given a subamplitude $\mathcal{A}_{n',k'}(I')$, in the following we will omit the dependence of the sub-amplitudes on $n'$ and we will use $\ldots$ for some or all indices in $I'$. They can be easily inferred from the context. All indices will be cyclically ordered $i<i+1<j<j+1<k<k+1$.

\begin{enumerate}
\item The \emph{Two-mass easy box} $\mathcal{C}^E_{ij}$ is a maximal cut with the following on-shell conditions:
\begin{equation}\label{2MEB}
\langle \mathcal{L} i-1,i\rangle=\langle \mathcal{L} i,i+1\rangle=\langle \mathcal{L} j-1,j\rangle=\langle \mathcal{L} j,j+1\rangle=0.
\end{equation}
There are two possible on-shell diagrams contributing to this cut, whose Leading Singularities are:
\begin{eqnarray}
\mathrm{LeS} \left[ \mathcal{A}_{-1}(i) \otimes \mathcal{A}_{0}(\ldots) \otimes \mathcal{A}_{-1}(j) \otimes \mathcal{A}_{1}(\ldots) \right]&=&\mathcal{A}_{NMHV}(j,\ldots,i),\\
\mathrm{LeS} \left[ \mathcal{A}_{-1}(i) \otimes \mathcal{A}_{1}(\ldots) \otimes \mathcal{A}_{-1}(j) \otimes \mathcal{A}_{0}(\ldots) \right]&=&\mathcal{A}_{NMHV}(i,\ldots,j).
\end{eqnarray}
\item The \emph{two-mass hard box} $\mathcal{C}^H_{ij}$ is a maximal cut with the following on-shell conditions:
\begin{equation}\label{2MHB}
\langle \mathcal{L} i-1,i\rangle=\langle \mathcal{L} i,i+1\rangle=\langle \mathcal{L} i+1,i+2\rangle=\langle \mathcal{L} j,j+1\rangle=0.
\end{equation}
There are two possible on-shell diagrams contributing to this cut, whose Leading Singularities are:
\begin{eqnarray}
\mathrm{LeS} \left[ \mathcal{A}_{-1}(i) \otimes \mathcal{A}_{0}(i+1) \otimes \mathcal{A}_{0}(\ldots, j) \otimes \mathcal{A}_{0}(\ldots) \right]&=& R_{i,i+1,i+2,j,j+1},\\
\mathrm{LeS} \left[ \mathcal{A}_{0}(i) \otimes \mathcal{A}_{-1}(i+1) \otimes \mathcal{A}_{0}(\ldots, j) \otimes \mathcal{A}_{0}(\ldots) \right]&=& R_{i-1,i,i+1,j,j+1}.
\end{eqnarray}
\item The \emph{three-mass box} $\mathcal{C}_{ijk}$ is a maximal cut with the following on-shell conditions:
\begin{equation}\label{3MB}
\langle \mathcal{L} i-1,i\rangle=\langle \mathcal{L} i,i+1\rangle=\langle \mathcal{L} j,j+1\rangle=\langle \mathcal{L} k,k+1\rangle=0.
\end{equation}
There is only one on-shell diagrams contributing to this cut, whose Leading Singularity is:
\begin{equation}
\mathrm{LeS} \left[ \mathcal{A}_{-1}(i) \otimes \mathcal{A}_{0}(\ldots,j) \otimes \mathcal{A}_{0}(\ldots, k) \otimes \mathcal{A}_{0}(\ldots) \right]= R_{i,j,j+1,k,k+1}.
\end{equation}
\end{enumerate}

\paragraph{Leading Singularities for N$^2$MHV 1 loop.}
For N$^2$MHV at one-loop, we have all cuts of the type appearing at NMHV, and in addition the \emph{four-mass box cut} appears from 8 points. 
This is associated to Leading Singularities which contains non-rational Yangian invariants, and, by Landau analysis, to algebraic singularities of the loop amplitude. We leave these cases for explorations in future works. 

N$^2$MHV Leading Singularities are in general expressed as:
\begin{equation}\label{yangianasR}
R_I \cdot R_J, \, \,  R_I \cdot \mathcal{A}_{NMHV}(J), \, \, \varphi R_I \cdot R_J
\end{equation}
where $I,J$ are lists of twistors (in general, expressed as intersection of lines or planes defined from $z_i$), $R_I$ are R-invariants with twistors in the list $I$, and $\varphi$ is an extra function, not relevant for our purposes.
Nevertheless, as discussed in \eqref{eq:LeSSumYI} all of them are just combinations of N$^2$MHV  Yangian invariants.
For the purpose of the paper, we are not interested in representations of Leading Singularities like \eqref{yangianasR}, but we will focus on their underline geometry. In particular, we are interested on the full list $\lbrace \mathcal{Y}_a \rbrace$ of Yangian invariants which can be used to express a given Leading singularity. In Section \ref{sec:cluster-adjacency-at-one-loop}, we will explain a way to obtain such list directly from the geometry of the loop amplituhedron.

\subsection{Landau Singularities from the amplituhedron}
\label{sec:LaS}
We will briefly review  how the Landau analysis can be used to infer singularities of the \emph{integral}, from the poles of the \emph{integrand}. First, we will review the original definition in terms of Feymann diagrams and then following \cite{Prlina:2017azl} we will review the role ampltiuhedron plays in this analysis.

\paragraph{Landau Singularities}

The concept of Landau singularities was originally introduced in 1959, when Landau stated a set of equations, called \emph{Landau Equations}, whose solutions parametrises the locus in the space of kinematic data where a given Feynman integral has branch points \cite{Landau:1959fi}.

Given a Feynman integral $I$ contributing to an $L$ loop  scattering amplitude in $D$ spacetime dimensions, we can always bring it to the following form by using Feynman parametrisation:

\begin{equation} \label{feynman}
I=c \int \prod_{a=1}^L d^D \ell_a \int_{\Delta_{\nu-1}}\frac{\mathcal{N}\left( \lbrace \ell_a \rbrace, \lbrace p_r \rbrace \right)}{\mathcal{D}^\nu(\alpha; \lbrace q_i \rbrace)}, \quad \mathcal{D}=\sum_{i=1}^{\nu} \alpha_i (q_i^2-m_i^2),
\end{equation}
where $c$ is just constant which does not enter our analysis, the integration is performed over the simplex $\Delta_{\nu-1}$, i.e. $\alpha_1+\ldots+\alpha_{\nu}=1$ and $\alpha_i \geq 0$, $q_i$ is the momentum flowing along the corresponding propagator $i$, $\lbrace p_r \rbrace$ are the momenta of external particles, and $\mathcal{N}$ is a function of the kinematic data.
It is known that the physical amplitude from \eqref{feynman} is obtained  by performing the integral over a particular contour defined by the $i \epsilon$ prescription in the propagators. However, in order to understand the analytic continuation outside the physical sheet in the space of kinematic, one has to study arbitrary contours.

The Landau analysis establishes that the integral $I$ in \eqref{feynman} \emph{can} develop singularities in either of the two following cases.
\begin{equation}\label{eq:LandauEq}
\sum_{i \in \mbox{\tiny{loop}}} \alpha_i q_i=0, \, \mbox{ for all loops} \quad \mbox{OR} \quad \alpha_i (q_i^2-m_i^2)=0, \, \forall i.
\end{equation}
In order to capture the analytic structure of $I$ away from the
physical sheet, one allows solutions of the following equations with
$\alpha_i$ and $\ell_a$ away from the physical contour as well.
When some of the $\alpha_i$ are different than zero, the second second case in Eq. \eqref{eq:LandauEq} corresponds to putting some internal propagators on-shell, and these will be related to `cuts'. 
In the following we will be interested only when solutions exist on codimension-one subspaces of the external kinematic space, i.e. when they are parametrised by the vanishing locus of a certain function of external kinematic.

We notice that the power of this method seems to be affected by two major inconveniences. Firstly, this analysis does not know about the numerator $\mathcal{N}$ in Eq.\eqref{feynman}, which can change the structure of singularities of the denominator, or even cancel some of them. Secondly, even when the numerator does not affect the singularities, 
singularities of individual Feynman integrals might not
survive the summation to remain singularities of the full amplitude.
In summary, the Landau analysis, even if predicts all potential singularities of the amplitude, in general it predicts many `spurious' singularities as well, which are not actual singularities of the amplitude.

In \cite{Dennen:2016mdk}, it was suggested that one can circumvent these issues by directly appealing to the geometry of the amplituhedron.

\paragraph{Landau Singularities form the loop amplituhedron} 
Given a Landau singularity corresponding to setting to zero a certain number of internal propagators, i.e. a \emph{cut}, this is an actual singularity of the amplitude if the cut corresponds to a boundary of the loop amplituhedron.

In order to make the connection with the amplituhedron more explicitly, as shown in \cite{Prlina:2017azl}, one can re-write the Landau equation in momentum twistors.
If a cut $\mathcal{C}$ is a collection of constraints of the type:
\begin{equation}\label{eq:Landaucut}
f_j(\mathcal{L},z)=0,
\end{equation}
where $\mathcal{L}$ collectively denotes the momentum twistors associated to loops $\mathcal{L}^{(1)},\ldots,\mathcal{L}^{(L)}$, and $\lbrace z \rbrace$ are momentum twistors encoding the kinematic data of external particles.
Then the Landau equations for this set of on-shell constraints include the above equations together with a set of equations of the type:
\begin{equation}
\sum_{j=1}^d \alpha_j \frac{\partial f_j(\mathcal{L}(\beta),z)}{\partial \beta_s}=0, \quad s=1,\ldots, 4L,
\end{equation}
where the $\beta$'s are $4L$ coordinates used to parametrise $\mathcal{L}^{(1)},\ldots,\mathcal{L}^{(L)}$. This latter
equations are often referred to as \emph{Kirchhoff conditions}.  We
observe that the Landau equations are $d+4L$ equations in $d+4L-1$
variables (since we can always rescale all the $\alpha$'s in the
Kirchhoff equations).  Therefore, one might expect that they do not
admit solutions for general kinematics.  For the purpose of this
analysis, one is then interested in knowing what the codimension-one loci
in kinematic space of $z$'s are, for which Landau
equations admit solutions (with $\alpha$'s not all zero).  If we
parametrise such loci as the vanishing set of the following function
\begin{equation}
\mathrm{LaS}[\mathcal{C},\mathcal{L}^*](z)=\prod_{t=1}^N a_{t}(z)=0, 
\end{equation} 
where $\mathcal{C}$ is the cut associated to the Landau Equations \eqref{eq:Landaucut}, $\mathcal{L}^*$ is one branch of solutions of the on-shell conditions we are considering, and $a_t(z)$ are certain polynomials of Plucker coordinates of $z$.
In the following, we will refer to $\mathrm{LaS}[\mathcal{C},\mathcal{L}^*]$ as the \emph{Landau singularity} associated to the cut $\mathcal{C}$ in the branch $\mathcal{L}^*$. With a slight abuse of terminology, we will also refer to $a_1,\ldots,a_N$ as corresponding Landau singularities.

Finally, a given Landau singularity $\mathrm{LaS}[\mathcal{C},\mathcal{L}^*]$ is a true singularity of the amplitude $\mathcal{A}^{(L)}_{n,k}$ if the loop amplituhedron  has a boundary of the type $\mathcal{B}[\mathcal{C},\mathcal{L}^*]$ \cite{Dennen:2016mdk}.

In summary, on one hand, the Landau analysis can connect the geometry of boundaries of the amplituhedron to the location of singularities of integrated amplitudes. On the other, the amplituhedron can tell which are the true singularities of the integrand, and therefore select the true Landau singularities, among the spurious ones coming from summing over Feynmann diagrams.

\subsubsection{Landau Singularities at 1 loop}
We report below the Landau singularities corresponding to some maximal cuts that will be relevant for our analysis. These can be found in \cite{Prlina:2017azl}, Table 1. We also report the points where the loop momenta localises on different cuts. In particular, for a maximal-cut $\mathcal{C}$, there are 2 solutions (one is parity conjugate to each other) $\mathcal{L}_1^*, \mathcal{L}^*_2$ each of which can be expressed in term of momentum twistors of external kinematic as:
\begin{equation} \label{eq:looploc}
\mathcal{L}^*_a=D[\mathcal{C},\mathcal{L}^*_a] \cdot z,
\end{equation}
 where $D[\mathcal{C},\mathcal{L}^*_a]$ is a $2 \times n$ matrix depending on Pluckers of twistors of external kinematics and $z$ is the $n \times 4$ matrix whose rows are $z_i$. In the following, only the non-zero columns of $D$ will be displayed explicitly. Moreover, we consider cyclically ordered indices $i<i+1<j<j+1<k<k+1$.

\begin{enumerate}

\item The \emph{two-mass hard box cut} $\mathcal{C}^E_{ij}$ in equation \eqref{2MEB} has in general 2 solutions:
\begin{equation}
\mathcal{L}^*_1=(ij), \quad \mathcal{L}^*_2=\bar{i} \cap \bar{j},
\end{equation}
the first is valid for $0 \leq k \leq n-6$ and the second for $2 \leq k \leq n-4$.
The corresponding matrices are\footnote{They are of course determined up to $GL(2)$ (and up to adding rows of $C$, see Def. \ref{def:loopAH}).}
\begin{equation}
D[{\mathcal{C}^E_{ij}},\mathcal{L}^*_1]= \begin{blockarray}{cc}
{\scriptstyle i} & {\scriptstyle j}  \\
\begin{block}{(cc)}
  1 & 0 \\
  0 & 1 \\
\end{block}
\end{blockarray}, \quad 
D[{\mathcal{C}^E_{ij}},\mathcal{L}^*_2]= \begin{blockarray}{ccc}
{\scriptstyle i-1 }& {\scriptstyle i} & {\scriptstyle i+1}  \\
\begin{block}{(ccc)}
 \langle i \bar{j} \rangle & -\langle i-1, \bar{j} \rangle & 0\\
 0& -\langle i+1, \bar{j} \rangle & \langle i \bar{j} \rangle \\
\end{block}
\end{blockarray}.
\end{equation}
For this cut and both of the branches\footnote{In general, we can have different Landau singularities for different branches of the same cut. However, this does not happen at one loop \cite{Prlina:2017azl}.} we have the following Landau singularities:
\begin{equation}
\mathrm{LaS}[\mathcal{C}^E_{ij},\mathcal{L}^*_1](z)=\mathrm{LaS}[\mathcal{C}^E_{ij},\mathcal{L}^*_2](z)=\langle i \bar{j} \rangle \langle \bar{i} j \rangle.
\end{equation}

\item The \emph{two-mass easy box cut} $\mathcal{C}^H_{ij}$ in equation \eqref{2MHB} has in general 2 solutions:
\begin{equation}
\mathcal{L}^*_1=\overline{i+1} \cap (ij j+1), \quad \mathcal{L}^*_2=\overline{i} \cap (i+1,j j+1)
\end{equation}
They are both valid for $1 \leq k \leq n-5$.
The corresponding matrices are:
\begin{eqnarray}
D[{\mathcal{C}^H_{ij}},\mathcal{L}^*_1] &=& \begin{blockarray}{ccc}
{\scriptstyle i} & {\scriptstyle i+1} & {\scriptstyle i+2}  \\
\begin{block}{(ccc)}
  1 & 0 & 0 \\
  0 & -\langle i,i+2,j,j+1 \rangle & \langle i,i+1,j,j+1 \rangle \\
\end{block}
\end{blockarray} \\
D[{\mathcal{C}^H_{ij}},\mathcal{L}^*_2] &=& \begin{blockarray}{ccc} 
{\scriptstyle i-1} & {\scriptstyle i} & {\scriptstyle i+1}  \\
\begin{block}{(ccc)}
  0 & 0 & 1 \\
  -\langle i,i+1,j,j+1 \rangle & \langle i-1,i+1,j,j+1 \rangle & 0 \\
\end{block}\scriptstyle 
\end{blockarray} \label{ex:Dmat2MHB}
\end{eqnarray}
For this cut we have the following Landau singularities:
\begin{equation}\label{ex:LaS2MH2}
\mathrm{LaS}[\mathcal{C}^H_{ij},\mathcal{L}^*_1](z)=\mathrm{LaS}[\mathcal{C}^H_{ij},\mathcal{L}^*_2](z)=\langle i, i+1, j,j+1 \rangle.
\end{equation}
\item The \emph{three-mass easy box} $\mathcal{C}_{ijk}$ in equation \eqref{3MB} has in general 2 solutions:
\begin{equation}
\mathcal{L}^*_1=(ijj+1)\cap (ik k+1), \quad \mathcal{L}^*_2=(\bar{i} \cap (jj+1),\bar{i} \cap (k k+1)).
\end{equation}
The first is valid for $1 \leq k \leq n-6$ and the second for $2 \leq k \leq n-5$.
The corresponding matrices are:
\begin{eqnarray}
D[\mathcal{C}_{ijk},\mathcal{L}^*_1] &=& \begin{blockarray}{ccc}
{\scriptstyle i} & {\scriptstyle i+1} & {\scriptstyle j}  \\
\begin{block}{(ccc)}
  1 & 0 & 0 \\
  0 & \langle i,j,k,k+1 \rangle & -\langle i,j+1,k,k+1 \rangle \\
\end{block}
\end{blockarray} \\
D[\mathcal{C}_{ijk},\mathcal{L}^*_2]&=& \begin{blockarray}{cccc}
{\scriptstyle j} & {\scriptstyle j+1} & {\scriptstyle k} & {\scriptstyle k+1}  \\
\begin{block}{(cccc)}
  -\langle \bar{i}, j+1 \rangle & \langle \bar{i}, j \rangle & 0 & 0 \\
  0 & 0 & -\langle \bar{i}, k+1 \rangle & \langle \bar{i}, k \rangle \\
\end{block}
\end{blockarray}.
\end{eqnarray}
For this cut we have the following Landau singularities:
\begin{equation}
\mathrm{LaS}[\mathcal{C}_{ijk},\mathcal{L}^*_1](z)=\mathrm{LaS}[\mathcal{C}_{ijk},\mathcal{L}^*_2](z)=\langle i (i-1,i+1) (j,j+1)(k,k+1)\rangle\, ,
\end{equation}
Our notation for twistor brackets throughout the paper follows closely
the literature, eg \cite{Prlina:2017azl}.
\end{enumerate}

\section{Cluster Adjacency in Yangian Invariants}
\label{sec:clust-adjac-yang}
We start unpacking cluster phenomena in the context of singularities of scattering amplitudes in $\mathcal{N}=4$ SYM.
In this section we focus on tree-level singularities, i.e. on poles of Yangian invariants (see Section \ref{sec:LeSAH}). 

In \cite{Drummond:2018dfd}, it was conjectured that, given a Yangian
invariant $\mathcal{Y}$ appearing in a BCFW representation of the
tree-level amplitude $\mathcal{A}_{n,k}$, then
\emph{all} its poles are \emph{cluster adjacent}, i.e. they are given
by some collection of $\mathcal{A}$-coordinates of the
$\operatorname{Gr}(4,n)$ cluster algebra that can be found together in
common cluster. In \cite{Mago:2019waa}, the conjecture was generalised
for all \emph{rational} Yangian invariants of $\mathcal{N}=4$ SYM, where rationality
in this context means intersection number one in the terminology of
\cite{ArkaniHamed:2012nw, Bourjaily:2012gy}.

In particular, in \cite{Drummond:2018dfd} cluster adjacency between
Yangian invariants was checked up to 8-points N$^2$MHV by looking at
Yangian invariants appearing into a specific representation of the
amplitude. This is not an exhaustive check since, starting from
8-points N$^2$MHV, one in general finds Yangian invariants which are
not related by cyclic symmetry to any of the Yangian invariants
appearing in a fixed representation.  Whereas in \cite{Mago:2019waa},
\emph{pair-wise} cluster adjacency between poles of Yangian invariants
was checked for all $k \leq 2$ and many $k=3$ Yangian invariants, by
employing Sklyanin Poisson brackets. However, this check obviously
does not imply cluster adjacency for the whole set of poles of a given
Yangian invariant.

In this section we prove that all (rational) N$^2$MHV Yangian invariants with are cluster adjacent by
explicit calculation of the relevant clusters. Moreover, we
provide explicitly the list
of their actual poles in terms of polynomial in Pluckers of momentum twistors which are $\mathcal{A}$-coordinates of the
$\operatorname{Gr}(4,n)$ cluster algebra in the files \texttt{yik2n\_.m}.

\paragraph{Poles of Yangian Invariants from the amplituhedron}
All Yangian invariants for $k=2$ has been fully classified in \cite{ArkaniHamed:2012nw}. However, the advantage of presenting Yangian invariants written as products of $R$-invariants (with, in some cases, auxiliary multiplicative rational functions) is shadowed by a drawback. In this way, the actual poles of the Yangian invariant are not always exposed: their numerator might indeed cancel some poles in the denominator. Moreover, the relation between these poles and $\mathcal{A}$-cluster coordinates of $\operatorname{Gr}(4,n)$ is not manifest either.

We will now explain how to improve on both aspects using a geometric approach from the amplituhedron, which automatically detects only actual poles of a Yangian invariant. Moreover, given an $\mathcal{A}$-cluster coordinate, it can easily tell whether it is a pole of a given Yangian invariant, without  computing its full expression.

We recall from section \ref{sec:LeSAH} that, given a $4k$-dimensional cell $S$ of $\operatorname{Gr}_+(k,n)$ with kinematic support, the Yangian invariant $Y_S$ is its full dimensional image in the amplituhedron and ${\mathcal{Y}}_S$ is the corresponding function of (super-)momentum twistors obtained from the Grassmannian integral formula.
Then there is a bijection between (actual) poles of $\mathcal{Y}_S$ and codimension-one boundaries of $Y_S$. Geometrically, these boundaries are simply described as the image of some of the boundaries of the cell $S$ into the amplituhedron. In details, let $\partial S^{(i)}$ be one $(4k-1)$-dimensional cell in the boundary of the cell $S$ that has a full dimensional image\footnote{Even if a cell $S$ has kinematic support, in general it has some boundaries whose dimension might be dropped when mapping into the amplituhedron. One has to disregard such boundaries, as they do not contribute to the codimension-one boundary of $Y_S$.} (i.e. $4k-1$) into the amplituhedron. Then the image of $\partial S^{(i)}$ is a boundary of $Y_S$. Moreover, being of codimension one, this  boundary in the amplituhedron lies on an hyper-surface determined by the vanishing locus of a function
\begin{equation}
P \left(\langle Y Z_{i_1} \, Z_{i_2}\, Z_{i_3} \, Z_{i_4} \rangle \right)=0,
\end{equation} 
which depends on the brackets defined in equation \eqref{eq:brackets}. 
By projecting to momentum twistor space, the corresponding pole of $\mathcal{Y}_S$ will be at
\begin{equation}
P \left(\langle z_{i_1} z_{i_2} z_{i_3} z_{i_4} \rangle \right)=0.
\end{equation}

Vice-versa, if we start with a polynomial\footnote{We assume the polynomials are irreducible in the Plucker variables.} $P \left(\langle z_{i_1} z_{i_2} z_{i_3} z_{i_4} \rangle \right)$ in Pluckers, e.g. an $\mathcal{A}$-cluster coordinate, then we can claim it is a pole of a given Yangian invariant $\mathcal{Y}_S$ if
\begin{equation}
\left. P \left(\langle Y Z_{i_1} \, Z_{i_2}\, Z_{i_3} \, Z_{i_4} \rangle \right) \right|_{Y=C \cdot Z}=0, \quad \forall C \in \partial S^{(i)},
\end{equation} 
where we parametrised $Y$ as in definition \eqref{def:AH}, i.e. $Y=C \cdot Z$, with $C$ any representative of a point in $\partial S^{(i)}$.
In this way we can detect (actual) poles of every Yangian invariant purely from the geometry of the amplituhedron.
In order to handle positroid cells in the Grassmannian, we used the Mathematica package \texttt{positroids.m} \cite{Bourjaily:2012gy}.

\subsection{Cluster adjacency in all (rational) N$^2$MHV Yangian Invariants}
\label{sec:adjac-all-rati}
In this section we report all poles of each N$^2$MHV Yangian invariant explicitly expressed as $\mathcal{A}$-cluster coordinates and explore their cluster properties.

\paragraph{n=6}
There is only one Yangian invariant and all its poles correspond to
frozen variables $\{\langle i,i+1,i+2,i+3 \rangle \}_{i \in
  [6]}$. Therefore, this Yangian invariant satifies cluster adjacency
trivially.

In the following we will denote
$\langle i \rangle:= \langle i, i+1,i+2,i+3 \rangle$, which will be
frozen variables in the \Gr{4,n}cluster algebra. Note that the
definition of $\ang{i}$ is depends on $n$. Nevertheless we surpress
this information to avoid notational clutter.

\paragraph{n=7}
There are only 3 Yangian invariants $\{\mathcal{Y}_{a} \}_{a \in [3]}$, up to cyclic symmetry. 
Only the Yangian invariants number $2,3$ are of the new type which appear at $n=7$, whereas $\mathcal{Y}_1$ is just a relabelling of the type $n=6$.
We list here the poles of each of them in terms of polynomial of brackets of momentum twistors:
\begin{eqnarray*}
\partial \mathcal{Y}_{1} &=& \{ \langle 2 \bar{6} \rangle,\langle 2367 \rangle, \langle \bar{3} 7 \rangle, \langle 2 \rangle, \langle 3 \rangle, \langle 4 \rangle \}\\
\partial \mathcal{Y}_{2} &=& \{\langle 1 \bar{4} \rangle, \langle 5 (67)(12)(34) \rangle, \langle 3 (45)(67)(12) \rangle, \langle 1 \rangle, \langle \bar{4} 7 \rangle, \langle 2 \rangle, \langle 3 \rangle, \langle 4 \rangle \}\\
\partial \mathcal{Y}_{3} &=&\{\langle 3471 \rangle, \langle 7 (12) (34)(56)\rangle,\langle \bar{2} 7 \rangle, \langle 3467 \rangle, \langle 1 \rangle, \langle \bar{4} 7 \rangle, \langle 3 \rangle, \langle \bar{3} 7 \rangle, \langle 4 \rangle \}
\end{eqnarray*} 
As expected, all the poles are cluster variables of $\operatorname{Gr}(4,7)$ and we checked they are cluster adjacent. 

\paragraph{n=8}
There are 24 Yangian invariants up to cyclic symmetry.  4 of them
are of $n=6$ type, 14 are of $n=7$ type. There are only 6 new
types which appear for $n=8$. We provide a full list of these Yangian
invariants in an ancillary file, \texttt{yik2n8.m}, and the labels we
use to denote them below refer to this list.

In this case, we observe explicitly that writing $k=2$ Yangian invariants in terms of products of R-invariants might obscure the actual poles. Let us consider the Yangian invariant $\mathcal{Y}_{11}$ which can be written as:
\begin{equation}\label{eq:doubleR}
\mathcal{Y}_{11}= R_{12345} \cdot R_{678,(123)\cap (45),5}.
\end{equation}
If we write the poles explicitly, we can see that some factorise
\begin{equation}
\langle (123) \cap (45)567\rangle =\langle 1235\rangle \langle 4567\rangle
\end{equation}
where $\langle 1235 \rangle$ is also a pole of $R_{12345}$ and will therefore seem to appear as a double pole in \eqref{eq:doubleR}. 
Since we appeal purely to the geometry of the amplituhedron, we will only see the actual poles of the Yangian invariants, and in this case $\langle 1235 \rangle$ is not an actual pole. In the following, we will write Yangian invariants which show this phenomenon explicitly, otherwise we will write their poles as union of 10 poles of 2 R-invariants.
\begin{eqnarray}
\partial \mathcal{Y}_{12} &=& \partial R_{123,(45)\cap \bar{7},8} \cup \partial R_{45678}\\
\partial \mathcal{Y}_{15} &=& \partial R_{81234} \cup \partial R_{45678}\\
\partial \mathcal{Y}_{16} &=& \partial R_{1234,\bar{5}\cap (78)} \cup \partial R_{45678}\\
\partial \mathcal{Y}_{11} &=& \{\langle 1 \bar{4}\rangle ,\langle 1245\rangle ,\langle 123 (45) \cap \bar{7}\rangle , \langle 1 \rangle, \langle 4578\rangle , \langle 2 \rangle, \langle \bar{5} 8\rangle, \langle 4 \rangle, \langle 5 \rangle \}\\
\partial \mathcal{Y}_{13} &=& \langle 6(13)(45)(78)\rangle, \langle 6(12)(45)(78) \rangle, \langle 123, \bar{5} \cap (78)\rangle, \langle 4 \bar{7}\rangle, \langle 123 (45)\cap \bar{7}\rangle,\\ && \langle \bar{5} 8\rangle, \langle 6(23)(45)(78), \langle 4 \rangle, \langle 5 \rangle \} \nonumber \\
  \partial \mathcal{Y}_{24} &=&
                                \lbrace\begin{aligned}[t]& \langle 1(34)(56)(78) \rangle, \langle 7 \rangle, \langle 6(12)(34)(78) \rangle, \langle 3(12)(56)(78) \rangle, \langle 1 \rangle, \langle 8(12)(34)(56) \rangle, \nonumber \\
                                  &\langle 5(12)(34)(78) \rangle, \langle 3 \rangle, \langle 2(34)(56)(78) \rangle, \langle 7(12)(34)(56) \rangle,\langle 5 \rangle, \langle 4(12)(56)(78) \rangle \rbrace \nonumber\end{aligned}\\ \label{eq:Y24}
\end{eqnarray}

All of these Yangian invariants, including those of
four-mass-box type, have poles that are polynomial in momentum twistor
brackets. These polynomials are all cluster variables of
$\operatorname{Gr}(4,8)$.

We verified that cluster variables corresponding to all the poles of
these Yangian invariants are cluster adjacent with a
single exception. Namely the Yangian invariant: $\mathcal{Y}_{24}$,
which corresponds to the four-mass box, contains non cluster-adjacent
poles. We will comment on this in section \ref{sec:four-mass-box}.

The \Gr{4,n} cluster algebras are infinite for $n\geq 8$ but recent
understanding \cite{Drummond:2019cxm,Arkani-Hamed:2019rds,Henke:2019hve} suggest natural truncations of these in terms of
\emph{positive tropical Grassmannians} or their generalisations. For $n=8$ there have
been three such constructions, by considering all tropicalised
Pl\"ucker coordinates, only a parity-invariant subset thereof, or the
parity completion of the set of Pl\"ucker coordinates. These, as polytopes, have 274,
260 and 548 vertices, respectively.

One may also wonder if and which of the proposed truncations of the
infinite cluster algebras via tropical fans do accommodate the adjacencies we found for $8$ points N$^2$MHV yangian invariants. We find that all Yangian invariants, except
$\mathcal{Y}_{24}$, are also cluster-adjacent in the more restrictive
sense of tropical fans. In particular, the corresponing g-vectors of
their unfrozen poles always form a cone of the tropical fan with 274
vertices, obtained by tropicalising the maximal parity-invariant
subset of the Pl\"ucker coordinates.

\paragraph{n=9}
There are 108 Yangian invariants up to cyclic symmetry. 10 of them
are of $n=6$ type, 56 are of $n=7$ type, and 38 are of $n=8$ type.
There are only 4 new types which appear for $n=9$.  We report below
the boundaries in terms of brackets of momentum twistors for these 4
types:
\begin{subequations}
\begin{eqnarray}
\partial \mathcal{Y}_{45} &=&\partial R_{12349} \cup \partial R_{56789}\\
\partial \mathcal{Y}_{46} &=&\partial R_{1234, (567)\cap (89)} \cup \partial R_{56789}\\
\partial \mathcal{Y}_{48} &=&\partial R_{1234, (56)\cap (789)} \cup \partial R_{56789}\\
  \partial \mathcal{Y}_{101}&=&\begin{aligned}[t]
    \lbrace
&\ang{5\,6\,\bar{2}\,\cap\,\bar{8}},
\ang{4\,6\,\bar{2}\,\cap\,\bar{8}},
\ang{4\,5\,\bar{2}\,\cap\,\bar{8}},\\
&  \ang{2\,3\,\bar{5}\,\cap\,\bar{8}},
\ang{1\,3\,\bar{5}\,\cap\,\bar{8}},
\ang{1\,2\,\bar{5}\,\cap\,\bar{8}},\\
&
\ang{8\,9\,\bar{2}\,\cap\,\bar{5}},
\ang{7\,9\,\bar{2}\,\cap\,\bar{5}},
\ang{7\,8\,\bar{2}\,\cap\,\bar{5}}\,
\rbrace \label{Y101}
    \end{aligned}\, ,
\end{eqnarray}
\end{subequations}
where the labels refer to the list in the ancillary file
\texttt{yik2n9.m}, where we give a full list of these Yangian
invariants (up to cyclic rotations). All 108 of these objects are
cluster adjacent in \Gr{4,9}, except the two, which are of the $n=8$
four-mass box type.

We also present the cluster that contains the poles of a Yangian
invariant which is particularly interesting, namely
$\mathcal{Y}_{101}$ in \eqref{Y101}. Informally, it is called the
\emph{spurion}, since it does not contain any physical
pole. Therefore, it can not appear in any of the BCFW representations
of the amplitude $\mathcal{A}_{9,2}$. Nevertheless, from the perspective of the
amplituhedron, it is a generalised triangle and can be part of a triangulation, giving a
representation of $\mathcal{A}_{9,2}$ not obtainable with standard BCFW.
\begin{figure}\centering
  
\begin{tikzpicture}
  \tikzstyle{frozen} = [draw=blue,fill=none,outer sep=2mm]
  \foreach \ni in {0,...,8}{
    \coordinate (n\ni) at (30+40*\ni:2);
    \pgfmathsetmacro{\nl}{int(Mod(\ni+1,9)+1)};
    \node (a\ni)  at (n\ni)  {$a_{\nl}$};
  }

  \foreach \ni in {0,...,8}{
    \coordinate (n\ni) at (30+20+40*\ni:3.7);
    \pgfmathsetmacro{\nl}{int(\ni+1)};
    \node[frozen] (f\ni) at (n\ni) {$f_{\nl}$};
  }

  \foreach \ni in {0,...,2}{
    \coordinate (n\ni) at (30+120*\ni:3.3);
    \pgfmathsetmacro{\nl}{int(\ni+1)};
    \node (s\ni) at (n\ni) {\color{gray}$s_{\nl}$};
  }

  \foreach \ni in {0,...,8}{
    \draw[-latex] let \n1={int(mod(\ni+1,9))} in (a\ni) -- (a\n1);
  }

  \foreach \ni in {0,...,8}{
    \draw[-latex]   (f\ni) -- (a\ni);    
  }
  \foreach \ni in {3,4,6,7,9,1}{
    \draw[-latex] let \n1={int(mod(\ni+1,9))} in let \n0={int(mod(\ni,9))} in (a\n1) -- (f\n0);
  }

  \foreach \ni in {1,2,3}{
    \draw[-latex] let \n3={int(mod(3*\ni,9))} in let \n1={int(mod(\ni,3))} in (s\n1) -- (f\n3);
    \draw[-latex] let \n3={int(mod(3*\ni,9))} in let \n1={int(mod(\ni,3))} in(a\n3) -- (s\n1);
    \draw[-latex] let \n3={int(mod(3*\ni-1,9))} in let \n1={int(mod(\ni,3))} in (s\n1) -- (f\n3);
  }\,
\end{tikzpicture}    
\caption{Quiver diagram for a \Gr{4,9} showing cluster adjacency of the ples of $\mathcal{Y}_{101}$ for $n=9$.}
\label{fig:flower}
  \end{figure}
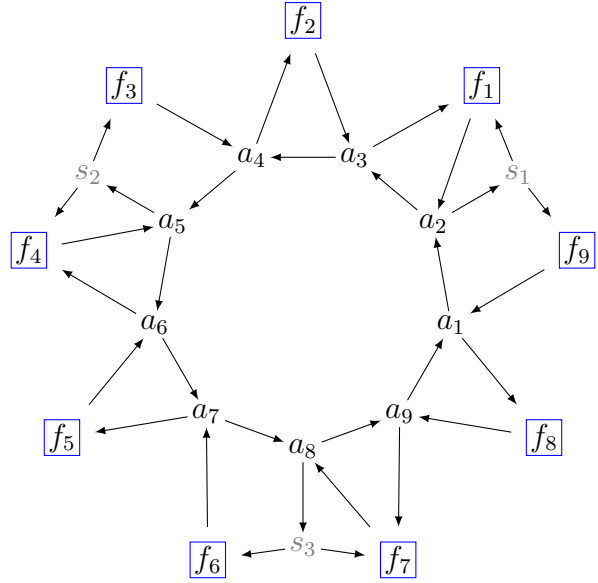

  This cluster has the quiver diagram displayed in figure
  \ref{fig:flower}, where we abbreviated the relevant
  $\mathcal{A}$-coordinates as
\begin{align}
a_1 &= \ang{5\,6\,\bar{2}\,\cap\,\bar{8}}&
a_2 &= \ang{4\,6\,\bar{2}\,\cap\,\bar{8}}&
a_3 &= \ang{4\,5\,\bar{2}\,\cap\,\bar{8}}\nonumber \\
a_4 &= \ang{2\,3\,\bar{5}\,\cap\,\bar{8}}&
a_5 &= \ang{1\,3\,\bar{5}\,\cap\,\bar{8}}&
a_6 &= \ang{1\,2\,\bar{5}\,\cap\,\bar{8}}\nonumber \\
a_7 &= \ang{8\,9\,\bar{2}\,\cap\,\bar{5}}&
a_8 &= \ang{7\,9\,\bar{2}\,\cap\,\bar{5}}&
a_9 &= \ang{7\,8\,\bar{2}\,\cap\,\bar{5}}\,,
\end{align}
while the remaning irrelevant ones are
\begin{align}
s_1 &= \ang{1456}&
s_2 &= \ang{1237}&
s_3 &= \ang{4789}\,.
\end{align}
These three irrelevant $\mathcal{A}$ coordinates can be freely mutated
and therefore one may say that this Yangian invariant corresponds to a
cube in the cluster polytope. Note also the $\mathbb{Z}_3$ symmetry of
the spurion is reflected in this cluster.

\begin{figure}
  \centering
  \centering
  \begin{tikzpicture}
  \tikzstyle{frozen} = [draw=blue,fill=none,outer sep=2mm];

  \foreach \col in {1,...,5}{
    \foreach \row in {1,...,3}{
      \pgfmathsetmacro{\ycord}{3-(1.5)*\row+1};
      \pgfmathsetmacro{\xcord}{2*\col+1};
      \coordinate (n\col\row) at  (\xcord,\ycord);

      }
    }

    \coordinate (nf1) at (2, 3.5);
    \coordinate (nf2) at (2, -1.5);
    \coordinate (nf6) at (12, -1.5);
    \coordinate (nf7) at (12, 3.5);

    \coordinate (nf3) at (4, -1.5);
    \coordinate (nf4) at (6, -1.5);
    \coordinate (nf5) at (8, -1.5);

    \coordinate (nf8) at (10, 3.5);
    \coordinate (nf9) at (8, 3.5);
    \coordinate (nf10) at (6, 3.5);

    \node (t1) at (n11) {$\ang{1\,2\,3\,5}$};
    \node (m1) at (n12) {$\ang{1\,2\,4\,5}$};
    \node (b1) at (n13) {$\ang{1\,3\,4\,5}$};

    \node (t2) at (n21) {$\ang{1\,2\,3\,6}$};
    \node (m2) at (n22) {$\ang{1\,2\,5\,6}$};
    \node (b2) at (n23) {$\ang{1\,4\,5\,6}$};

    \node (t3) at (n31) {$\ang{1\,2\,6\,10}$};
    \node (m3) at (n32) {$\ang{1\,2\,6\,7}$};
    \node (b3) at (n33) {$\ang{1\,5\,6\,7}$};

    \node (t4) at (n41) {$\ang{1\,6\,9\,10}$};
    \node (m4) at (n42) {$\ang{1\,6\,7\,10}$};
    \node (b4) at (n43) {$\ang{1\,6\,7\,8}$};

    \node (t5) at (n51) {$\ang{6\,8\,9\,10}$};
    \node (m5) at (n52) {$\ang{6\,7\,9\,10}$};
    \node (b5) at (n53) {$\ang{6\,7\,8\,10}$};

    \node[frozen] (f1) at (nf1) {$\ang{1\,2\,3\,4}$};
    \node[frozen] (f2) at (nf2) {$\ang{2\,3\,4\,5}$};
    \node[frozen] (f3) at (nf3) {$\ang{3\,4\,5\,6}$};
    \node[frozen] (f4) at (nf4) {$\ang{4\,5\,6\,7}$};
    \node[frozen] (f5) at (nf5) {$\ang{5\,6\,7\,8}$};
    \node[frozen] (f6) at (nf6) {$\ang{6\,7\,8\,9}$};
    \node[frozen] (f7) at (nf7) {$\ang{7\,8\,9\,10}$};
    \node[frozen] (f8) at (nf8) {$\ang{1\,8\,9\,10}$};
    \node[frozen] (f9) at (nf9) {$\ang{1\,2\,9\,10}$};
    \node[frozen] (f10) at (nf10) {$\ang{1\,2\,3\,10}$};

    \draw[-latex] (t1) -- (t2);
    \draw[-latex] (t2) -- (t3);
    \draw[-latex] (t3) -- (t4);
    \draw[-latex] (t5) -- (t4);

    \draw[-latex] (m1) -- (m2);
    \draw[-latex] (m2) -- (m3);
    \draw[-latex] (m4) -- (m3);
    \draw[-latex] (m5) -- (m4);

    \draw[-latex] (b1) -- (b2);
    \draw[-latex] (b3) -- (b2);
    \draw[-latex] (b4) -- (b3);
    \draw[-latex] (b5) -- (b4);

    \draw[-latex] (t1) -- (m1);
    \draw[-latex] (m1) -- (b1);
    \draw[-latex] (t2) -- (m2);
    \draw[-latex] (m2) -- (b2);
    \draw[-latex] (m3) -- (t3);
    \draw[-latex] (m3) -- (b3);
    \draw[-latex] (m4) -- (t4);
    \draw[-latex] (b4) -- (m4);
    \draw[-latex] (m5) -- (t5);
    \draw[-latex] (b5) -- (m5);
    
    \draw[-latex] (b2) -- (m1);
    \draw[-latex] (b3) -- (m2);
    \draw[-latex] (m2) -- (t1);
    \draw[-latex] (t3) -- (m4);
    \draw[-latex] (m4) -- (b5);
    \draw[-latex] (t4) -- (m5);

    \draw[-latex] (t2) -- (f10);
    \draw[-latex] (f10) -- (t3);
    \draw[-latex] (t3) -- (f9);
    \draw[-latex] (f9) -- (t4);
    \draw[-latex] (t4) -- (f8);
    \draw[-latex] (f8) -- (t5);

    \draw[-latex] (b1) -- (f3);
    \draw[-latex] (f3) -- (b2);
    \draw[-latex] (b2) -- (f4);
    \draw[-latex] (f4) -- (b3);
    \draw[-latex] (b3) -- (f5);
    \draw[-latex] (f5) -- (b4);

    \draw[-latex] (f1) -- (t1);
    \draw[-latex] (f2) -- (b1);
    \draw[-latex] (t5) -- (f7);
    \draw[-latex] (b5) -- (f6);    
\end{tikzpicture}\qquad.
\caption{The cluster in \Gr{4,10} demonstrating the adjacency of the Yangian invariant $\mathcal{Y}_1$ for $n=10$.}
\label{yangiann10}
\end{figure}
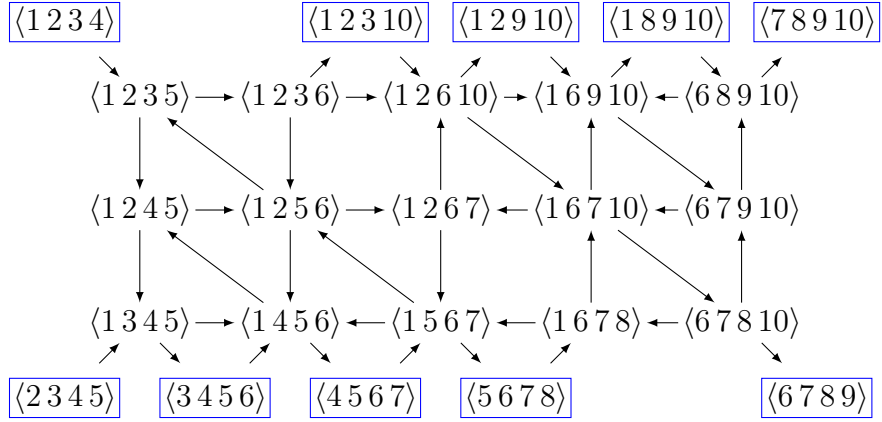

\paragraph{n=10}
There are 395 Yangian invariants up to cyclic symmetry.
22 of them are of $n=6$ type, 168 are of $n=7$ type,  174 are of $n=8$ type, 30 are of $n=9$ type and only 1 is of $n=10$.
As observed in \cite{ArkaniHamed:2012nw}, there are no new types of Yangian invariants beyond $n=10$. This comes immediately from the fact that Yangian invariants correspond to $4k$-dimensional cells in $\operatorname{Gr}_+(k,n)$, whose number of types is bounded for fixed $k$. 
The only new type of Yangian invariant for $n=10$ is very simple:
\begin{equation}
\partial \mathcal{Y}_{1}= \partial R_{12345} \cup \partial R_{6789,10}.
\end{equation}
In Fig. \ref{yangiann10}, we represent a cluster in $\operatorname{Gr}(4,10)$ cluster algebra which contains all poles of $\mathcal{Y}_{1}$. We observe that all poles of $R_{12345}$ are in the left-most position, whereas all the poles of $R_{6789,10}$ are in the right-most position.

With this, we proved cluster adjacency for \emph{all} N$^2$MHV Yangian
Invariants corresponding to generalised triangles. Moreover, we
observe that Yangian invariants of the four-mass box type, which are
not generalised triangles, do not satisfy cluster adjacency. The
corresponding cluster algebras are infinite, and one may wonder how
one can check this conclusively. We comment on this in the next
section.

\subsection{Four-mass box Yangian invariants}
\label{sec:four-mass-box}
The $n=8$ Yangian invariant $\mathcal{Y}_{24}$ corresponds to a
four-mass box cut.  These types of Yangian invariants fall in the
category of Yangian invariants with intersection number higher than
one, see Section \ref{sec:LeSAH} and they have always been excluded in
cluster-adjacency analysis, e.g. in
\cite{Mago:2019waa}.

In particular, for the case of $\mathcal{Y}_{24}$, points in the
amplituhedron have 2 pre-images in the associated cell $S_{24}$ in
$\operatorname{Gr}_+(k,n)$.  From an algebraic perspective, this
corresponds to the fact that $\mathcal{Y}_{24}$ can be expressed as
the sum of two contributions:
\begin{equation}\label{eq:sumfourmass}
\mathcal{Y}_{24}=\mathcal{Y}^{(1)}_{24}+\mathcal{Y}^{(2)}_{24},
\end{equation}
each of which contains square-roots, however the sum is of course
rational. Moreover, the boundaries of $\mathcal{Y}_{24}$ will only
correspond to the actual poles of the sum in equation
\eqref{eq:sumfourmass}.  We reported these poles in
equation \eqref{eq:Y24} and checked that they are \emph{not} cluster
adjacent\footnote{$\mathcal{Y}_{24}$, as other Yangian invariants with
  intersection number higher than one, can be rewritten as a sum of
  other rational Yangian invariants, each of which we showed to
  satisfy cluster adjacency.}.  The infinite nature of the \Gr{4,8}
cluster algebra makes it difficult to perform an exhaustive check,
especially when the poles may not be mutation pairs\footnote{See
  Section \ref{sec:cluster-adjacency} for the terminology.}.
Nevertheless, we explain below how we can argue a conclusive check.

We can easily find a cluster containing only two of the poles of this
Yangian invariant, e.g.
\begin{equation*}
  \langle 6 (12)(34)(78)\rangle, \qquad \langle 8(12)(34)(56)\rangle \, .
\end{equation*}
Then, by freezing these two nodes and mutating in all other
directions, we can start exploring the ``face'' correspinding to these
and ask whether we are able to generate any of the other poles of this
Yangian invariant. This turns out to be infinite. However after a
relatively small number of mutations one finds that all mutations are
exhausted apart from those corresponding to the 1-dimensional infinite
sub affine-$A_2$ sequences. Each mutation in these sequences produces
to a Pl\"ucker polynomial of an increasing degree, exhausting the
possibilities of generating any of the poles of equation \eqref{eq:Y24}.

\section{Patterns in Leading and Landau Singularities}
\label{sec:cluster-adjacency-at-one-loop}
In this section we state the main conjecture of this work. We enhance
the tree-level cluster adjacency of Yangian invariants explored in
Section \ref{sec:clust-adjac-yang}, to include information of
loop-level singularities, i.e. Landau singularities. In particular, we provide
evidence that all poles of a Yangian invariant in a given cut, and
the corresponding Landau singularity can be found together in a
cluster.

Cluster adjacency seems to know about compatibility between different
singularities, or equivalently, between boundaries. We have seen at
tree level how the collection of poles of a Yangian invariant (or
equivalently, of their boundaries) corresponds to cluster variables in
a common cluster. One can naturally extend this compatibility thinking
of a Yangian invariant as being located `inside' a given Leading
Singularity. Algebraically, this means it can be used as an addend to
express the Leading Singularity. Geometrically, this means literally
that the Yangian is inside the codimension-$4L$ boundary of the loop
amplituhedron which corresponds to the maximal cut giving the Leading
Singularity, as explained in Section \ref{sec:LeSAH}.  By Landau
analysis, we have seen how this boundary of the loop amplituhedron
(equivalently, the Leading Singularity) is accessed by the
\emph{integrated} amplitude having a branch points in the
corresponding Landau singularity.  Vice-versa, given a Landau
singularity which corresponds to branch points of the integrated
amplitude, by `reverse' Landau analysis we can list the maximal cuts
of the integrand which are responsible for these singularities. The
Leading Singularities of these maximal cuts will be then expressed in
terms of Yangian invariants, which themselves have certain
poles. Cluster algebras seem to tell us that we can find the Landau
singularity \emph{and} all the poles of a given Yangian as above in a
common cluster.

Let us state our conjecture more explicitly. 
Given a maximal-cut $\mathcal{C}$ of an $L$-loop $n$-point N$^k$MHV amplitude and a branch of its solutions $\mathcal{L}^*$, let $\mathrm{LaS}[\mathcal{C},\mathcal{L}^*]$ be the corresponding Landau singularity and let $\mathrm{LeS}[\mathcal{C},\mathcal{L}^*]$ be the corresponding Leading Singularity, as defined in sections \ref{sec:LeSAH} and \ref{sec:LaS}. Let us express them as:
\begin{equation}
\mathrm{LaS}[\mathcal{C},\mathcal{L}^*](z)=\prod_{t=1}^N a_{t}(z),\quad
\mathrm{LeS}[\mathcal{C},\mathcal{L}^*]=\sum_i \mathcal{Y}_i,
\end{equation}
where $\lbrace a_1(z),\ldots,a_N(z) \rbrace$ and the poles of the $n$-points N$^k$MHV Yangian invariants $\mathcal{Y}_i$
are $\mathcal{A}$-coordinates of the $\operatorname{Gr}(4,n)$ cluster algebra. Then we conjecture:

\begin{conj}\label{conj:strong}
$\lbrace \mathcal{Y}_i, a_1,\ldots,a_N \rbrace$ are \emph{cluster adjacent}, i.e. all the poles of $\mathcal{Y}_i$ and $a_1,\ldots,a_N$ can be found in a common cluster of the $\operatorname{Gr}(4,n)$ cluster algebra.
\end{conj}
 This refers to \emph{any} Yangian $\mathcal{Y}$ invariant that can be used to represent the given Leading Singularity $\mathrm{LeS}$.

 We will refer to the cluster adjacency predicted by these conjectures
 as the \emph{LL-Cluster Adjacency} (i.e. Leading and Landau
 singularities Cluster Adjacency). In this paper we checked for all
 amplitudes up to 1 loop and 9-points (both NMHV and N$^2$MHV)
 that LL-cluster adjacency \eqref{conj:strong} holds
 true. We leave the checks for higher loops or points for future work.

\paragraph{Yangian Invariants from the loop amplituhedron.}

For an $L$-loop, $n$-point N$^k$MHV amplitude, let $\mathcal{C}$ be a maximal-cut and $\mathcal{L}^*$ one branch of its solutions. Then we can directly use the definition of the loop amplituhedron to determine whether a given Yangian invariant $\mathcal{Y}$ can be used to express the corresponding Leading Singularity $\mathrm{LeS}[\mathcal{C},\mathcal{L}^*]$.
In geometric terms this means, that the Yangian invariant $Y$ is inside the codimension-$4L$ boundary $\mathcal{B}[\mathcal{C},\mathcal{L}^*]$ of the loop amplituhedron $\mathcal{A}^{(L)}_{n,k}$.
 
As mentioned in equation  \eqref{eq:looploc}, on the maximal-cut $\mathcal{C}$ and on the branch of solutions $\mathcal{L}^*$, the loop momentum twistors are localised in terms of twistors of external kinematic:
\begin{equation}\label{eq:locloop}
{\mathcal{L}^{(l)}}^*={D^{(l)}}^*(\langle z_{i_1}z_{i_2}z_{i_3}z_{i_4} \rangle)\cdot z, \quad l=1,\ldots,L,
\end{equation}
 where ${D^{(l)}}^*$ are $2 \times n$ matrices depending on Pluckers of the twistors of external kinematics and $z$ is the $n \times 4$ matrix whose rows are $z_i$.
 
On the amplituhedron side, on the boundary $\mathcal{B}[\mathcal{C},\mathcal{L}^*]$, the loop momentum twistors are localised as:
\begin{equation}
{\mathcal{L}^{(l)}}^*={D^{(l)}}^*(\langle Y Z_{i_1}Z_{i_2}Z_{i_3}Z_{i_4} \rangle)\cdot Z, \quad l=1,\ldots,L,
\end{equation}
where ${D^{(l)}}^*$ are the same as in equation  \eqref{eq:locloop}, however their dependence on $\langle z_{i_1}z_{i_2}z_{i_3}z_{i_4} \rangle$ has been uplifted in the ampliutedron into a dependence on $\langle Y Z_{i_1}Z_{i_2}Z_{i_3}Z_{i_4} \rangle$.

Let $S$ be a $4k$-dimensional cell in $\operatorname{Gr}_+(k,n)$ with kinematic support.
Then a Yangian invariant $Y_S$ belongs to the boundary $\mathcal{B}[\mathcal{C},\mathcal{L}^*]$ if the positivity conditions in \eqref{eq:poscond} are satisfied, i.e.
\begin{equation}\label{eq:loclpos}
\begin{pmatrix}
C \\
{D^{(i_1)}}^*|_{Y=C \cdot Z} \\
\vdots \\
{D^{(i_s)}|_{Y=C \cdot Z}}^*
\end{pmatrix}
\end{equation}
are totally positive matrices, with $0\leq s \leq L$, for all representative matrices $C$ in the cell $S$. In \eqref{eq:loclpos}, we denoted ${D^{(i_a)}}^*|_{Y=C \cdot Z}$ as the matrix which depends on $\langle Y Z_{i_1}Z_{i_2}Z_{i_3}Z_{i_4} \rangle$, with $Y$ in the image of the cell $S$, i.e. $Y=C \cdot Z$.
As in previous sections, in order to handle positroid cells in the positive Grassmannian, we use the Mathematica package \texttt{positroids.m}.

Using this procedure, by scanning over all $4k$-dimensional cells (with kinematic support) in $\operatorname{Gr}_+(k,n)$, we get a list $\lbrace S_i \rbrace_i$ such that $\lbrace Y_{S_i} \rbrace_i$ are all the Yangian invariants in the boundary $\mathcal{B}[\mathcal{C},\mathcal{L}^*]$. Finally, this means that we obtain the list of Yangian invariants $\lbrace \mathcal{Y}_{S_i} \rbrace_i$ which can appear as summands to represent the Leading Singularity $\mathrm{LeS}[\mathcal{C},\mathcal{L}^*]$.

\subsection{LL-Cluster Adjacency for 1 loop NMHV}
\label{sec:LLNMHV}
Let us consider the case of one-loop $n$-points NMHV amplitudes and state the expected LL-cluster adjacencies by matching Yangian invariants in representations of a Leading Singularity with the corresponding Landau singularities associated to the same maximal-cut.
We note that our studies focus on the non-trivial cases when Landau singularities are not only product of frozen variables, which are the ones presented in Section \ref{sec:LaS}. 
 We will refer to Section \ref{sec:LeSAH} for the corresponding Leading Singularities. 
 
The Landau singularity for the two easy-mass box cut $\mathcal{C}^E_{ij}$ is given by the product of the cluster variables $\langle i \bar{j} \rangle $ and $\langle \bar{i} j \rangle $. Whereas the Leading Singularities for the cut $\mathcal{C}^E_{ij}$ are $\mathcal{A}_{\mathrm{ NMHV}}(i,\ldots,j)$ and $\mathcal{A}_{\mathrm{NMHV}}(j,\ldots,i)$. It is straightforward to see that the R-invariants which can appear in a representation of these Leading Singularities are just the ones of the type $R_I$, with $I$ a 5-element subset of the set $\lbrace j,\ldots, i\rbrace$ or of $\lbrace i,\ldots, j\rbrace$. 
The two hard-mass box cut $\mathcal{C}^H_{ij}$ has a Landau singularity which is just the cluster variable $\langle i,i+1,j,j+1 \rangle$, whereas its Leading Singularities are $R_{i-1,i,i+1,j,j+1}$ and $R_{i,i+1,i+2,j,j+1}$. Since $\langle i,i+1,j,j+1 \rangle$ is already a pole of both the latter two R-invariants, LL-cluster adjacency is trivially satisfied in this case. Finally, the three-mass box $\mathcal{C}_{ijk}$ has an associated Landau singularity which is $\langle i (i-1 \, i+1) (j j+1) (k k+1)\rangle$ and the Leading Singularity is $R_{i,j,j+1,k,k+1}$.

In summary, LL-cluster adjacency for all points one-loop NMHV rely on the following conjecture:

\begin{conj}\label{conj:adjalln}
In the $\operatorname{Gr}(4,n)$ cluster algebra, there are always clusters containing the following lists of $\mathcal{A}$-coordinates together:
\begin{eqnarray}
 \{ \partial R_{I}, \langle i \bar{j} \rangle, \langle \bar{i} j \rangle \}, \quad I \in {\{j,\ldots,i\} \choose 5},{\{i,\ldots,j\} \choose 5} \\
  \{\partial  R_{i,j,j+1,k,k+1}, \langle i (i-1 \, i+1) (j j+1) (k k+1)\rangle \}, \label{conj:adjall2}
\end{eqnarray}
where by $\partial R_{J}$ we denoted the list of all poles of the R-invariant $R_J$, and $i<i+1<j<j+1<k<k+1$ are cyclically ordered indices in $\lbrace 1, \ldots, n \rbrace$.
\end{conj}

While we provide an \emph{all $n$ proof} of the second type of adjacencies \eqref{conj:adjall2} in \ref{proof:alln}, in the following we report checks of LL-cluster adjacencies up to 9-points, with corresponding cluster mutations, as explained in Appendix \ref{app:mutations}.
\subsubsection{LL Cluster Adjacency for 1 loop 7 points NMHV}  
Up to cyclic shift, for $n=1$ there only 3 types of R-invariants: 
\begin{equation}
(12),(13),(14), \quad (i_1,\ldots,i_{n-5}):=R_{[n]/\{i_1,\ldots,i_{n-5}\}}
\end{equation}
The adjacencies between Landau and Leading Singularities reads:
\begin{eqnarray*}
(12) &\mbox{ is CA with }& \langle 3 \bar{7} \rangle, \langle \bar{3} 7 \rangle \\
(13) &\mbox{ is CA with }& \langle 2(13)(45)(67)\rangle\,.
\end{eqnarray*}
Both adjacencies are manitested in the cluster polytope by the
presence of a subpolytope made out of clusters that contain all active
poles of the Yangian invariants and the Landau singularities.  The
R-invariant $(12)$ has three active poles, and together with the
Landau singularity, these correspond to four
$\mathcal{A}$-coordinates. The remaining two degrees of freedom
correspond to a pentagonal face of the cluster polytope, ie an $A_2$
subalgebra.

The pentagons that correspond to $\bigl\{(12), \ang{1367}\bigr\}$ and
$\bigl\{(12), \ang{2347}\bigr\}$ share an edge, ie the subpolytope of
two cluster that contain the poles of $(12)$ as well as both parity
conjugate Landau singularities. This will be a recurring feature for
higher $n$: When the Landau analysis predicts the product of two
parity conjugate Pl\"ucker coordinates, there is a cluster that
contains both of them as well as the Yangian invariant. We will omit
the parity conjugate singularity to keep notation short.

The R-invariant $(13)$ has 4 active poles, and together with the
Landau singularity $\ang{2(13)(45)(67)}$, this adjacency corresponds
to a line segment.

\subsubsection{LL Cluster Adjacency for 1 loop 8 points NMHV}  
There are 7 cyclically inequivalent R-invariants in this and we find
that all associated Landau singularities are cluster adjacent. In
Table \ref{tab:n8nmhv} we provide the checks for the cases that are
not implied by the $n=7$ case. We also omit the cases where the Landau
singularity is a pole of the NMHV invariant.

\begin{table}
  \centering
  \begin{tabular}{llll}
  $R$-Inv.& Landau sing.&Mutation sequence & Subalgebra\\
  \midrule
  {(123)} & $\ang{3 \bar{8}}\ang{\bar{3}8}$&(1,5,9,1,2,8,1,2,5,7,1,2,4,2,3,1,5,6)&$D_5$\\
  &$\ang{4 \bar{1}}\ang{1\bar{4}}$&(1,4,7,5,9,1,5,8,1,4,5)&$D_4$\\
  \midrule
  {(124)} & $\ang{3 \bar{8}}\ang{8 \bar{3}}$&(2,6,5,8,9,4,5,4,7,3,1,2,4)&$A_3$\\
  \midrule
  {(125)} & $\ang{3 \bar{8}}\ang{8 \bar{3}}$&(2,5,8,6,9,1,2,4,5,4,7,3,1,4)&$A_1\times A_1$\\
  \midrule
  {(126)} & $\ang{3 \bar{8}}\ang{\bar{3} 8}$&(2,5,8,2,4,6,2,4,5,9,1,2,1,7,1,3)&$A_1\times A_1$\\
  \midrule
  {(127)} & $\ang{3 \bar{8}}\ang{\bar{3} 8}$&(2,5,8,5,7,9,6,1,2,5,1,2,4,1,2,3)&$A_3$\\
  &$\ang{ 8(71)(34)(56)}$&(7,8,9,2,6,1,2,5,4,7,1,2)&$D_4$\\
  \midrule
  {(147)} & $\ang{ 8(71)(23)(56)}$&(2,3,6,7,8,9,2,5,4,7)&$A_1\times A_1 \times A_1$ 
\end{tabular}

\caption{Checks of cluster-adjacencies of $n=8$ $R$-invariants and NMHV Landau singularities. The mutation sequences describe a cluster starting from the initial cluster. See Appendix A for labelling conventions. The subalgebras donate the residual freedom of mutations, which leave the collective set of poles invariant.}
\label{tab:n8nmhv}
\end{table}

For example the cluster-adjacency statements
\begin{eqnarray}
(123) &\mbox{ is CA with }& \langle 4 \bar{8} \rangle, \langle \bar{4} 8 \rangle \\
(124) &\mbox{ is CA with }& \langle 3(24)(56)(78)\rangle
\end{eqnarray}
are implied by the adjacency of $(12)$ to $\ang{3\bar{7}}$ and the
ajacency of $(13)$ to $\ang{2(13)(45)(67)}$. We also omit everywhere
Landau singularities that are pole sof the Yangian invariants,
trivially satisfying cluster adjacency based on that of Yangian
invariants.

\subsubsection{LL Cluster Adjacency for 1 loop 9 points NMHV}  
There are 14 R-invariants up to cyclic symmetry. The adjacencies we
need to check along with the verifications are listed in Table
\ref{tab:n9nmhv}. We note that cluster adjacencies between Landau
singularities and Leading Singularities at $n=7$ and $n=8$ are
embedded in $n=9$, as we should expect.

\begin{table}
  \centering
  \begin{tabular}{llll}
  \toprule
  $R$-Inv.& Landau sing.&Mutation sequence & Subalgebra\\*
  \midrule
  {(1234)}    & $\ang{1 \bar{4}}\langle 4\bar{1} \rangle$&(4,7,10)&$E_7$\\
  \midrule
  {(1235)} & $\langle \bar{1} 4\rangle\ang{1 \bar{4}}$&  (2,1,4,4,10)&$E_6$\\
  & $\ang{3 \bar{9}} \ang{9 \bar{3}}$&(2,1,4,4,10,11,12)&$E_7$\\
  \midrule
  {(1236)} & $\ang{4 \bar{1}} \ang{1 \bar{4}}$&(3,6,9,10,11,12,1,2,1,3)&$E_7$\\
  & $\ang{3 \bar{9}} \ang{9 \bar{3}}$&(3,6,9,10,11,12,1,6,1)&$E_6$\\
  \midrule
  {(1237)} & $\ang{4 \bar{1}} \ang{1 \bar{4}}$&(7,1,2,1,3,6)&$E_6$\\  
  &$\ang{3 \bar{9}} \ang{9 \bar{3}}$&(7,1,3,1)&$E_6$\\
  \midrule
  {(1238)} & $\ang{4 \bar{1}} \ang{1 \bar{4}}$&(1,2,3,5,10)&$\tilde D_5$\\
  &$\ang{3 \bar{9}} \ang{9 \bar{3}}$&(1,2,3,5,11,7,10,12)&$E_6$\\
  & $\ang{4 \bar{9}} \ang{9 \bar{4}}$&(1,2,3,5,7,10)&$D_6$\\
  &$\langle 9(81)(45)(67)\rangle$&(1,2,3,5,6,5)&$\tilde{E}_6$\\*
  \midrule
  {(1278)} & $\ang{9 \bar{6}} \ang{6 \bar{9}}$&(5,1,2,4,7,8,5,10,2,5,2,7,8,10)&$E_6$\\
  & $\ang{3 \bar{9}} \ang{9 \bar{3}}$&(5,1,2,4,7,8,5,10,2,8)&$E_6$\\
  & $\langle 9(81)(34)(56)\rangle$&(5,1,2,4,7,8,5,10,2,5)&$E_7$\\
  \midrule
  {(1246)} & $\ang{3 \bar{9}} \ang{9 \bar{3}}$ &(2,5,7,8,1,3,6)&$A_2\times A_2$\\*
  \midrule
  {(1247)} & $\ang{3 \bar{9}} \ang{9 \bar{3}}$&(3,4,6,8,9,1,5,1)&$A_3\times A_2 \times A_1$\\
  &  $\langle 3(24)(56)(89)\rangle$&(3,4,6,8,9,1,6)&$D_6 \times A_1 $\\
  \midrule
  {(1248)} & $\ang{3 \bar{9}} \ang{9 \bar{3}}$ &(2,6,7,4,5,8,9,12,2,5,2,6,7,9)&$D_6 \times A_1$\\*
  \midrule
  {(1267)} & $\ang{8 \bar{5}} \ang{5 \bar{8}}$&(2,5,6,1,3,1,2)&$A_4\times A_1$\\  \midrule
  {(1257)} & $\ang{3 \bar{9}} \ang{9 \bar{3}}$&(3,6,7,8,9,1,6,1)&$A_3\times A_2 \times A_2$\\
  &  $\langle 6(57)(89)(34) \rangle$&(1,4,5,3,6,3)&$A_5\times A_1$\\
  \midrule
  {(1258)} & $\ang{3 \bar{9}} \ang{9 \bar{3}}$&(3,4,5,6,1)&$A_3\times A_1 \times A_1$\\
  \midrule
  {(1268)}&$\ang{3 \bar{9}} \ang{9 \bar{3}}$&(2,5,8,7,10,1,3)&$A_5$
\end{tabular}
  \caption{Checks of cluster-adjacencies of $n=9$ $R$-invariants and NMHV Landau singularities. The mutation sequences describe a cluster starting from the initial cluster. See Appendix A for labelling conventions. The subalgebras donate the residual freedom of mutations, which leave the collective set of poles invariant.}
\label{tab:n9nmhv}
\end{table}

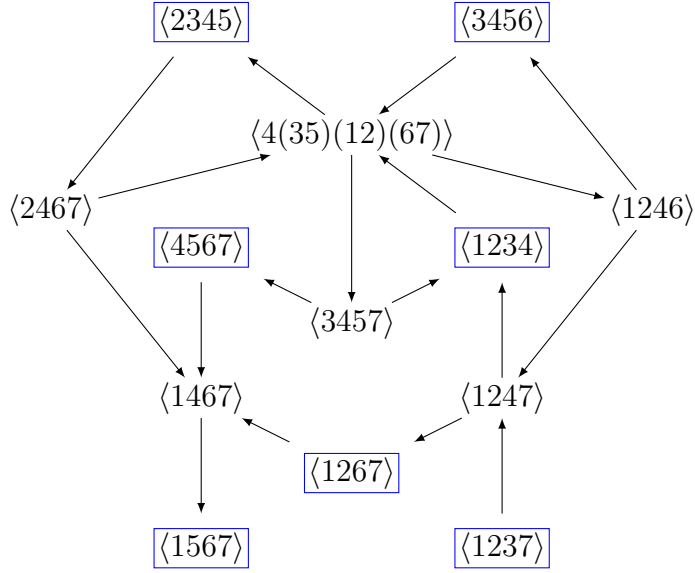
\begin{figure}
  \centering
  \begin{tikzpicture}
  \tikzstyle{frozen} = [draw=blue,fill=none,outer sep=2mm]
  \node[frozen] (f1) at (2,1) {$\ang{1234}$};
  \node[frozen] (f2) at (-2,4) {$\ang{2345}$};
  \node[frozen] (f3) at (+2,4) {$\ang{3456}$};
  \node[frozen] (f4) at (-2,1) {$\ang{4567}$};
  \node[frozen] (f5) at (-2,-3) {$\ang{1567}$};
  \node[frozen] (f6) at (0,-2) {$\ang{1267}$};
  \node[frozen] (f7) at (2,-3) {$\ang{1237}$};

  \node (n1) at (-4,1.5) {$\ang{2467}$};
  \node (n2) at (0,2.5) {$\ang{4(35)(12)(67)}$};
  \node (n3) at (4,1.5) {$\ang{1246}$};
  \node (n4) at (-2,-1) {$\ang{1467}$};
  \node (n5) at (2,-1) {$\ang{1247}$};
  \node (n6) at (0,0) {$\ang{3457}$};

  \draw[-latex] (f7) -- (n5);
  \draw[-latex] (n5) -- (f6);
  \draw[-latex] (f6) -- (n4);
  \draw[-latex] (n4) -- (f5);

  \draw[-latex] (n3) -- (f3);
  \draw[-latex] (f3) -- (n2);
  \draw[-latex] (n2) -- (f2);
  \draw[-latex] (f2) -- (n1);

  \draw[-latex] (n1) -- (n4);
  \draw[-latex] (n3) -- (n5);
  \draw[-latex] (n2) -- (n6);

  \draw[-latex] (n1) -- (n2);
  \draw[-latex] (n2) -- (n3);
  \draw[-latex] (n6) -- (f4);
  \draw[-latex] (n6) -- (f1);
  \draw[-latex] (f1) -- (n2);
  \draw[-latex] (f4) -- (n4);
  \draw[-latex] (n5) -- (f1);
\end{tikzpicture}\qquad.

\caption{The quiver diagram of a \Gr{4,7} cluster containing the poles of the $R$-invariant $R_{12467}$ and the $\mathcal{A}$-coordinate $\ang{ 4(23)(12)(67)}$. This cluster is enough to prove the LL cluster adjacency for all one-loop NMHV amplitudes.}
\label{fig:proofcluster}
\end{figure}

\subsubsection{An all-$n$ proof}
\label{proof:alln}
It is straightforward to prove that the Landau singularities
$\langle i\, (i-1 i+1)\, (j j+1)\,(k k+1 \rangle$ are
cluster-adjacent to the poles of the $R$-invariants
$R_{i\,j\,j+1\,k\,k+1}$ in the strict sense, ie there exists a
cluster that contains both this Landau singularity and all the poles
of the said $R$-invariant in any \Gr{4,n} cluster algebra, with $n$
sufficiently large to accommodate the former. Without loss of
generality, we can fix $j=1$ and assume $k+1<i-1$. All other cases are
related to this by cyclic symmetry.

We will show this by explicitly constructing such a cluster, closely
following \cite{Drummond:2018dfd} where the cluster-adjacency of any
$R$-invariant was proved based on partial rotations.

We first find a cluster that contains the poles of $R_{12467}$ and
$\langle 4(23)(12)(67) \rangle$ in the \Gr{4,7} cluster algebra. This
cluster can be obtained after a sequence of mutations, which we shall
dentote by $\Sigma_0$. The resulting cluster has the quiver
diagram displayed in figure \ref{fig:proofcluster}.

The cluster we aim to find is just a relabelling of the cluster above,
and this can be achieved through partial cyclic rotations. In
particular we need to find a sequence of rotations that maps the
labels of $(1,2,3,4,5,6)$ to $(k,k+1,i-1,i,i+1,1,2)$. These rotations
are
\begin{equation}
  (1,2,3,4,5,6,7)
  \xrightarrow{-2 \bigl|_{k+4}} 
  (k+3,k+4,1,2,3,4,5)
  \xrightarrow{-3 \bigl|_{i+1}}
  (k,k+1,i-1,i,i+1,1,2)\, ,
\end{equation}
where $r\bigl|_m$ denote $r$ rotations in the \Gr{4,m} algebra. In
\cite{Drummond:2018dfd} it was explained how to find a mutation sequence that
realises such a transformation, and we denote this sequence with
$\Sigma^r_m$. For negative $r$, it is understood that the mutation sequence is applied in reverse

If we then apply the mutation sequence $\Sigma_0$ to the appropriately
relabelled cluster, in other words, if we mutate the \Gr{4,n} initial
cluster in the sequence
\begin{equation}
  \Sigma^{-3}_{i+1} \, \Sigma^{-2}_{k+4}\, \Sigma_{0},
\end{equation}
we obtain a cluster which contains all the poles of the $R$-invariant
$R_{1\,2\, k\,k+1\, i}$ as well as the letter
$\ang{i(i-1\,i+1)(12)(k\,k+1)}$.

\subsubsection{An amplitude in a manifestly LL cluster-adjacent form} \label{sec:NMHV7explicit}
In \cite{Drummond:2018dfd}, it was argued that to make the adjacency
of the final entries of symbols with the R-invariants of NMHV loop
amplitudes, one has to expand the symbol of the amplitude over the full
set of R-invarants, which satisfy linear, six-term relations among
them. This way of writing the amplitude is certainly not unique due to
these identities, and cluster-adjaceny between the R-invariants and the
final entries can be made manifest in a number of ways.

We can suggest that our observation of the cluster adjacency of Landau
singularities and R-invariants as a further constraint on the final
entry condition. More precisely, we can rule out any final entry -
R-invariant pairs that are not related to each other through LL-cluster adjacency.

It turns out that, the 1-loop NMHV amplitude can be uniquely fixed
fixed in a form in which it obeys the adjacency discussed above:
\begin{equation}
  \begin{aligned}[t]
  \mathcal{A}^{\mathrm{BDS-like},(1)}_{1,7}=&-(13)\, \bigl[a_{11} \otimes a_{62} + a_{13} \otimes a_{62}\bigr] \\
 &+(14)\,\bigl[-a_{11} \otimes a_{11} + a_{11} \otimes a_{14} + a_{14} \otimes a_{11} - a_{14} \otimes a_{14}\bigr] \\
 &+(12)\,\bigl[
 \begin{aligned}[t]
 &a_{11} \otimes a_{15} - a_{11} \otimes a_{22} - a_{11} \otimes a_{31} + a_{12} \otimes a_{15} - a_{12} \otimes a_{22}\\
   &- a_{12} \otimes a_{31} - 2 a_{15} \otimes a_{15} + a_{15} \otimes a_{22} + a_{15} \otimes a_{31}\bigr]\,,
 \end{aligned}
  \end{aligned}
\end{equation}
where the notation $a_{ij}$ for the symbol letters follows the
literature on seven-point amplitudes.

It would be interesting to show that any NMHV 1-loop amplitude can be
written in such a way. Then one may speculate whether this leads to a
final entry condition for a given MHV degree and loop order.

\subsection{LL cluster adjacency for 1 loop N$^2$MHV}
\label{sec:LLN2MHV}
We now use results from Section \ref{sec:clust-adjac-yang}, in particular the lists of poles of (rational) N$^2$MHV Yangian invariants expressed as cluster coordinates, to test LL Cluster Adjacency at one-loop for N$^2$MHV amplitudes. Moreover, for a given cut, we use positivity as in equation  \eqref{eq:loclpos} to get the full list of N$^2$MHV Yangian invariants which appear in a representation of the corresponding Leading Singularity.
For example, if we consider the two-mass hard box cut $\mathcal{C}^H_{15}$ and the branch as in equation \eqref{ex:Dmat2MHB}, then for 8 points we find a list of six N$^2$MHV Yangian invariants, whereas for $n=9$ and the same cut and branch we find there are 21. Each of these N$^2$MHV Yangian invariants has poles which can be found in the same cluster together with the Landau singularity of the corresponding cut in equation  \eqref{ex:LaS2MH2}, i.e. $\langle 1256\rangle$.

\paragraph{n=7}
This case is related to $n=7$ NMHV case discussed above through parity
conjugation. Up to cyclic shift, for $n=7$ there only 3 types of
Yangian invariants $\mathcal{Y}_{1,2,3}$, which are parity conjugates
of $(12),(13),(14)$ respectively. Their poles have been explicitly
presented in Section \ref{sec:adjac-all-rati}.

For these Yangian invariants, we only have the following associated Landau
singularitiy that is not implied by a case worked out earlier in the paper:
\begin{longtable}{llll}
  Y-Inv.& Landau sing.&Mutation sequence & Subalgebra\\*
  \midrule
  \endhead
  $\mathcal{Y}_1$&$\langle 3 \bar{6} \rangle, \langle 6 \bar{3} \rangle$&(3)&$A_1$\\
  $\mathcal{Y}_2$&$\ang{4(12)(35)(67)}$ &(6,5,4,1,2)&$A_1$\, ,
\end{longtable}
\noindent where, as before, the mutation sequences describe a cluster proving
the cluster adjacency.

\paragraph{n=8}
There are 24 Yangian invariants up to cyclic symmetry. We see that
for some of the Landau singularity - Yangian invariant pairs, cluster
adjacency is implied either by checks for $n=6$ and $n=7$, or the by
the cluster adjacency of the Yangian invariant itself, when the Landau
singularity is a pole of the former. The remaining cases are listed in
the table below together with the clusters that contain of all said singularities:
\begin{longtable}{llll}
  Y-Inv.& Landau sing.&Mutation sequence & Subalgebra\\
  \midrule
  \endhead
  $\mathcal{Y}_1$&$\ang{1378}$ &(2,5,8,6,9,1,2,4,5,1,4,7)&$A_3\times A_2$\\
  $\mathcal{Y}_{10}$&$\ang{8(71)(23)(56)}$&(7,8,9,3,6,2,5,7,1,4)&$A_3\times A_1$\\
  $\mathcal{Y}_{14}$&$\ang{7(68)(23)(45)}$&(1,5,9,3,5,6,1,2,3,4,5,4,8,7)&$A_1\times A_1$\\
  $\mathcal{Y}_{18}$&$\ang{8(71)(23)(45)}$&(7,8,9,4,5,6,3,1,2,7,8,1,4,7)&$A_1\times A_1$
\end{longtable}

We have also checked the the cluster-adjacency properties for Landau
singularities and Yangian invariants with relevant for N$^2$MHV $n=9$
one-loop amplitudes. These are too lengthy to present here and we
refer to the ancillary file \texttt{LLCAk2n9.m}.

\section{Conclusions and Outlook} \label{sec:conl}
Cluster phenomena have become increasingly relevant in understanding singularities of scattering amplitudes. The remarkable observation that building blocks of scattering amplitudes in planar $\mathcal{N}=4$ SYM satisfy a property called \emph{cluster adjacency}, provided both paths to understand their deep mathematical structures and tools to perform computations which otherwise would be beyond reach.
At tree level, scattering amplitudes are rational functions whose singularity structure is encoded in the location of its poles. They can be expressed as sums of Yangian invariants, which have their own singularities, some of which are `spurious' as they do not appear in the final amplitude. Nevertheless, collections of all poles of each Yangian invariant which can appear in a representation of tree-level scattering amplitudes seem themselves to be part of a beautiful mathematical story.   

In this paper, we argued for an enhancement of the phenomenon of
cluster adjacency of Yangian invariants to include singularities of
loop amplitudes in $\mathcal{N}=4$ SYM. In particular, via an
amplituhedron-based approach we observed a new manifestation of
cluster adjacency for Leading and Landau singularities, that we called
``LL cluster adjacency'' for brevity.  Given a maximal-cut of a loop
amplitude, the corresponding Landau singularities are found in the
same cluster as the poles of each Yangian invariant which can appear
in a representation of the Leading Singularity related to the cut.
Moreover, we checked LL-cluster adjacencies for all one-loop
amplitudes, both NMHV and N$^2$MHV, up to 9 points.  Interestingly,
one-loop NMHV 7-points amplitude are uniquely fixed by LL cluster adjacencies, once these are interpreted as a
final-entry conditions.  On the way, we proved that all N$^2$MHV
Yangian invariants corresponding to generalised triangles are cluster
adjacent, confirming the conjectures of
\cite{Drummond:2018dfd,Mago:2019waa}. We also show that, for Yangian
invariants of the four-mass box type, the poles of the rational sum of
their algebraic terms violate cluster adjacency.

Studies of cluster adjacency of Yangian invariants
\cite{Lukowski:2019sxw,Mago:2019waa} motivate the question of whether
this phenomenon should be regarded as a built-in mathematical feauture
of Yangian invariants that is by their definition, or whether it is a
physical constraint. While cluster-adjacency may be a mathematical
fact for rational Yangian invariants, the inclusion of Landau
singularities is certainly a new, ``physical'' information, just like
extended-Steinmann conditions on symbol letters. Our observation calls
for a geometric understanding that unifies all the related
incarnations of cluster-adjacency and makes manifest how the more
physical ones are implied by the mathematical ones and vice versa.

At each loop $L$ and helicity sector $k$, LL cluster adjacencies
provide a natural set of pairings between Yangian invariants and
symbol letters of the corresponding Landau Singularities. This set is
in general much smaller than the full set of adjacencies obtained from
\Gr{4,n} cluster algebras, which is loop- and helicity- agnostic. It
would be interesting to see if LL cluster adjacencies could be
interpreted as \emph{refined} set one can use to contain amplitudes at
fixed loop order and MHV degree, instead of relying on the full
cluster algebra. We provided an example of how this principle can be
interprete as a final-entry condition and fix the $n=7$ one-loop
amplitude. Such a principle would be even more restrictive than the
recently proposed truncations of infinite cluster algebras
\cite{Drummond:2019cxm,Arkani-Hamed:2019rds,Henke:2019hve}, but
without further examples with higher multiplicity and loop order, it is merely a wishful speculation.

On this regard, extending our analysis to non-rational Yangian invariants and Landau singularities corresponding to algebraic letters could be natural direction to pursue. There have been many promising results on how to understand algebriac letters in a cluster algebra fashion with the help of tropical positive Grassmannians  \cite{Henke:2019hve,Drummond:2019cxm,Arkani-Hamed:2019rds}. However, a full understanding of \Gr{4,n} infinite cluster algebras is still missing, e.g. notions of `adjacencies' involving algebriac letters have still to be defined.

Our work is in the direction of making steps towards answering the long-standing question of how the cluster structure of integrands in $\mathcal{N}=4$ SYM theory is related to the cluster structure of the integrated amplitudes.
The manifestation of the physical information carried by LL cluster adjacencies which relates Leading Singularities with Landau Singularities shows further evidence that cluster phenomena know about the mathematical structure encoding (the singularities of) loop amplitudes.

\subsubsection*{Acknowledgements}
We are grateful to J. Drummond and M. Spradlin for valuable
discussions. M.P would also like to thank S. Stanojevic, N. Henke and G. Papathanasiou for stimulating discussions.
This project has received funding from the European
Research Council (ERC) under the European Union's Horizon 2020
research and innovation programme (grant agreement No. 724638).

\appendix
\section{Initial clusters and the encoding of mutations} \label{app:mutations}
To prove the cluster adjacency properties we claim the hold, we
present clusters containing these poles as the results of a sequence
of mutations starting from the initial cluster.

\begin{figure}[h]
  \centering
  \begin{tikzpicture}
  \tikzstyle{frozen} = [draw=blue,fill=none,outer sep=2mm, inner sep=1mm];

  \foreach \col in {1,...,4}{
    \foreach \row in {1,...,3}{
      \pgfmathsetmacro{\ycord}{3-(1.5)*\row+1};
      \pgfmathsetmacro{\xcord}{2*\col+1};
      \coordinate (n\col\row) at  (\xcord,\ycord);

      }
    }

    \coordinate (nf1) at (1,4);
    \foreach \col in {2,3,...,7}{
      \pgfmathsetmacro{\xcord}{2*\col-1};
      \coordinate (nf\col) at (\xcord, -2);
    }
    \foreach \row in {8,...,10}{
      \pgfmathsetmacro{\ycord}{2*\row-18};
      \coordinate (nf\row) at (14, \ycord);
    }

    \draw[line width=2mm, orange!40!white, ->] ($(n11)-(0.5,0.5)$)  -- ($(n13)-(0.5,0.75)$);
    \draw[line width=2mm, orange!40!white, ->] ($(n21)-(0.5,0.5)$)  -- ($(n23)-(0.5,0.75)$);

    \node[line width=2mm, orange!40!white, ->] at ($(n11)-(1,0.25)$)  {\bf\fontfamily{phv}\selectfont 1};
    \node[line width=2mm, orange!40!white, ->] at ($(n11)-(1,1.75)$)  {\bf\fontfamily{phv}\selectfont 2};
    \node[line width=2mm, orange!40!white, ->] at ($(n11)-(1,3.25)$)  {\bf\fontfamily{phv}\selectfont 3};
    \node[line width=2mm, orange!40!white, ->] at ($(n21)-(1,0.25)$)  {\bf\fontfamily{phv}\selectfont 4};
    \node[line width=2mm, orange!40!white, ->] at ($(n22)-(1,0.25)$)  {\bf\fontfamily{phv}\selectfont 5};
    \node[line width=2mm, orange!40!white, ->] at ($(n23)-(1,0.25)$)  {\bf\fontfamily{phv}\selectfont 6};

    \node (t1) at (n11) {$\ang{1\,2\,3\,5}$};
    \node (m1) at (n12) {$\ang{1\,2\,4\,5}$};
    \node (b1) at (n13) {$\ang{1\,3\,4\,5}$};

    \node (t2) at (n21) {$\ang{1\,2\,3\,6}$};
    \node (m2) at (n22) {$\ang{1\,2\,5\,6}$};
    \node (b2) at (n23) {$\ang{1\,4\,5\,6}$};

    \node (t3) at (n31) {$\dotsm$};
    \node (m3) at (n32) {$\dotsm$};
    \node (b3) at (n33) {$\dotsm$};
    \node (f4) at (nf4) {$\dotsm$};

    \node (t4) at (n41) {\color{white}\raisebox{5mm}{AAA}};
    \node (m4) at (n42) {\color{white}\raisebox{5mm}{AAA}};
    \node (b4) at (n43) {\color{white}\raisebox{5mm}{AAA}};
    
    \node[frozen,anchor=west] (t4c) at ($(n41)+(-0.5,0)$) {$\ang{1\,2\,3\,n}$};
    \node[frozen,anchor=west] (m4c) at ($(n42)+(-0.5,0)$) {$\ang{1\,2\,n-1\,n}$};
    \node[frozen,anchor=west] (b4c) at ($(n43)+(-0.5,0)$) {$\ang{1\,n-2\,n-1\,n}$};

    \node[frozen] (f1) at (nf1) {$\ang{1\,2\,3\,4}$};
    \node[frozen] (f2) at (nf2) {$\ang{2\,3\,4\,5}$};
    \node[frozen] (f3) at (nf3) {$\ang{3\,4\,5\,6}$};
    \node[] (f4) at (nf4) {\color{white}\raisebox{5mm}{AAA}};
    \node[] (f5) at (nf5) {\color{white}\raisebox{5mm}{AAA}};
    \node[frozen,anchor=west] (f5c) at ($(nf5)+(-0.5,0)$) {$\ang{n-3\,n-2\,n-1\,n}$};
    \raisebox{5mm}{AAA}

    \draw[-latex] (t1) -- (t2);
    \draw[-latex] (t2) -- (t3);
    \draw[-latex] (t3) -- (t4);

    \draw[-latex] (m1) -- (m2);
    \draw[-latex] (m2) -- (m3);
    \draw[-latex] (m4) -- (m3);

    \draw[-latex] (b1) -- (b2);
    \draw[-latex] (b3) -- (b2);
    \draw[-latex] (b4) -- (b3);

    \draw[-latex] (t1) -- (m1);
    \draw[-latex] (m1) -- (b1);
    \draw[-latex] (t2) -- (m2);
    \draw[-latex] (m2) -- (b2);
    \draw[-latex] (m3) -- (t3);
    \draw[-latex] (m3) -- (b3);

    \draw[-latex] (b2) -- (m1);
    \draw[-latex] (b3) -- (m2);
    \draw[-latex] (m2) -- (t1);
    \draw[-latex] (m4) -- (t3);
    \draw[-latex] (b4) -- (m3);
    \draw[-latex] (m3) -- (t2);

    \draw[-latex] (f1) -- (t1);
    \draw[-latex] (b1) -- (f2);
    \draw[-latex] (b2) -- (f3);
    \draw[-latex] (f3) -- (b1);
    \draw[-latex] (f4) -- (b2);
    \draw[-latex] (f5) -- (b3);

\end{tikzpicture}\qquad.

\caption{Initial cluster for \Gr{4,n} and a numbering of its nodes to encode mutation sequences}
\label{fig:initial}
\end{figure}
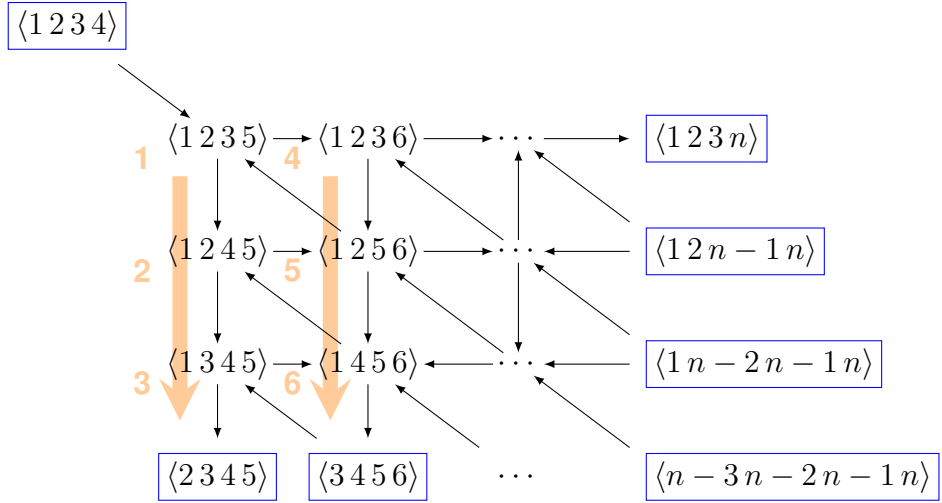

We enumerate the active nodes of the initial cluster, starting from
$\ang{1235}$ going downwards and continuing in the second column
starting with $\ang{1236}$, numbered 4.  In figure \ref{fig:initial}
we remind the reader the initial cluster for \Gr{4,n} and describe
this way of labelling the nodes. Clearly the mutation sequence that
relates two clusters is not unique.

\bibliographystyle{JHEP}
\bibliography{biblio}

\end{document}